\documentclass[prd,aps,twocolumn,preprintnumbers, showpacs, nofootinbib,superscriptaddress,notitlepage]{revtex4-1}
\usepackage{amssymb,amsthm,amsmath}
\usepackage{graphicx}   
\usepackage{color}      
\usepackage{slashed}    
\usepackage{verbatim}
\usepackage[normalem]{ulem}
\usepackage{rotating}   
\usepackage{multirow}   

\begin{document}

\title{Nonperturbative Determination of Collins-Soper Kernel from  Quasi Transverse-Momentum Dependent Wave Functions}

\collaboration{\bf{Lattice Parton Collaboration ($\rm {\bf LPC}$)}}

\author{Min-Huan Chu}
\affiliation{INPAC, Key Laboratory for Particle Astrophysics and Cosmology (MOE),  Shanghai Key Laboratory for Particle Physics and Cosmology, School of Physics and Astronomy, Shanghai Jiao Tong University, Shanghai 200240, China}

\affiliation{Yang Yuanqing Scientific Computering Center, 
Tsung-Dao Lee Institute, Shanghai Jiao Tong University, Shanghai 200240, China}

\author{Zhi-Fu Deng}
\affiliation{INPAC, Key Laboratory for Particle Astrophysics and Cosmology (MOE),  Shanghai Key Laboratory for Particle Physics and Cosmology, School of Physics and Astronomy, Shanghai Jiao Tong University, Shanghai 200240, China}

\author{Jun Hua}
\affiliation{Guangdong Provincial Key Laboratory of Nuclear Science, Institute of Quantum Matter, South China Normal University, Guangzhou 510006, China}
\affiliation{Guangdong-Hong Kong Joint Laboratory of Quantum Matter, Southern Nuclear Science Computing Center, South China Normal University, Guangzhou 510006, China}

\affiliation{INPAC, Key Laboratory for Particle Astrophysics and Cosmology (MOE),  Shanghai Key Laboratory for Particle Physics and Cosmology, School of Physics and Astronomy, Shanghai Jiao Tong University, Shanghai 200240, China}

\author{Xiangdong Ji}
\affiliation{Department of Physics, University of Maryland, College Park, MD 20742, USA}
\affiliation{Center for Nuclear Femtography,
1201 New York Ave., NW, Washington DC, 20005, USA}

\author{Andreas Sch\"afer}
\affiliation{Institut f\"ur Theoretische Physik, Universit\"at Regensburg, D-93040 Regensburg, Germany}

\author{Yushan Su}
\affiliation{Department of Physics, University of Maryland, College Park, MD 20742, USA}

\author{Peng Sun}
\affiliation{Nanjing Normal University, Nanjing, Jiangsu, 210023, China}

\author{Wei Wang}
\email{Corresponding author:wei.wang@sjtu.edu.cn}
\affiliation{INPAC, Key Laboratory for Particle Astrophysics and Cosmology (MOE),  Shanghai Key Laboratory for Particle Physics and Cosmology, School of Physics and Astronomy, Shanghai Jiao Tong University, Shanghai 200240, China}

\author{Yi-Bo Yang}
\affiliation{CAS Key Laboratory of Theoretical Physics, Institute of Theoretical Physics, Chinese Academy of Sciences, Beijing 100190, China}
\affiliation{School of Fundamental Physics and Mathematical Sciences, Hangzhou Institute for Advanced Study, UCAS, Hangzhou 310024, China}
\affiliation{International Centre for Theoretical Physics Asia-Pacific, Beijing/Hangzhou, China}
\affiliation{School of Physical Sciences, University of Chinese Academy of Sciences,
Beijing 100049, China}

\author{Jun Zeng}
\affiliation{INPAC, Key Laboratory for Particle Astrophysics and Cosmology (MOE),  Shanghai Key Laboratory for Particle Physics and Cosmology, School of Physics and Astronomy, Shanghai Jiao Tong University, Shanghai 200240, China}

\author{Jialu Zhang}
\affiliation{INPAC, Key Laboratory for Particle Astrophysics and Cosmology (MOE),  Shanghai Key Laboratory for Particle Physics and Cosmology, School of Physics and Astronomy, Shanghai Jiao Tong University, Shanghai 200240, China}

\author{Jian-Hui Zhang} 
\affiliation{Center of Advanced Quantum Studies, Department of Physics, Beijing Normal University, Beijing 100875, China}

\author{Qi-An Zhang}
\email{Corresponding author:zhangqa@buaa.edu.cn}
\affiliation{
School of Physics, Beihang University, Beijing 102206, China}
\affiliation{Key Laboratory for Particle Astrophysics and Cosmology (MOE),
Shanghai Key Laboratory for Particle Physics and Cosmology,
Tsung-Dao Lee Institute, Shanghai Jiao Tong University, Shanghai 200240, China}

\begin{abstract}
  In the framework of large-momentum effective theory at one-loop matching accuracy, we perform a lattice calculation of the Collins-Soper kernel which governs the rapidity evolution of  transverse-momentum-dependent (TMD) distributions. We first obtain 
  the quasi TMD wave functions at three different meson momenta 
  on a lattice with valence clover quarks on a dynamical HISQ sea 
  and lattice spacing $a=0.12$~fm from MILC, and renormalize the pertinent linear divergences using Wilson loops. Through one-loop matching to the light-cone wave functions, we determine the Collins-Soper kernel with transverse separation up to 0.6~fm.  We study the systematic uncertainties from operator mixing and scale dependence, as well as the impact from higher power corrections. Our results potentially allow for a determination of the soft function and other transverse-momentum dependent quantities at one-loop accuracy. 
\end{abstract}

\maketitle
\section{Introduction}

Understanding the internal, three-dimensional structure of hadrons, such as the proton, is an important goal in nuclear and particle physics. In this regard, the transverse momentum-dependent (TMD) parton distribution functions (TMDPDFs)~\cite{Collins:1981uk,Collins:1981va} play an important role as they characterize their intrinsic transverse 
partonic structure. These distributions are also essential ingredients in the description of multi-scale and non-inclusive processes, such as Drell-Yan production of electroweak gauge bosons or Higgs bosons or semi-inclusive deep-inelastic scattering with small transverse momentum, in the context of QCD factorization theorems. As a result, they have received considerable attention in the past few decades (for a review, see Ref.~\cite{Angeles-Martinez:2015sea}). More accurate experimental measurements are expected in the coming decades from JLab 12 GeV~\cite{Dudek:2012vr} and the Electron-Ion Colliders in the US~\cite{Accardi:2012qut,AbdulKhalek:2021gbh} and in China~\cite{Anderle:2021wcy}.

In contrast to the TMDPDFs that encode the probability density of parton momenta in hadrons, the transverse momentum-dependent wave functions (TMDWFs) offer a probability amplitude description of the partonic structure of hadrons, from which one can potentially calculate various quark/gluon distributions. In the QCD factorization involving transverse momentum, they are the most important ingredients to predict physical observables in exclusive processes, for instance, weak decays of heavy $B$ meson~\cite{Keum:2000wi,Lu:2000em} which are valuable to extract the CKM matrix element, and to probe new physics beyond the standard model. However, due to the lack of  knowledge of TMDWFs, the one-dimensional lightcone distribution amplitudes (LCDAs) are used instead in most analyses of $B$ decays \cite{Keum:2000wi,Lu:2000em,Nagashima:2002ia}, resulting in uncontrollable errors.  The unprecedented  precision of experimental measurements  of $B$ decays~\cite{Cerri:2018ypt} urgently requires a reliable theoretical  knowledge of TMDWFs. 

A common feature of TMDPDFs and TMDWFs is that they depend both on the longitudinal momentum fraction $x$ and on the transverse spatial separation of partons.  Considerable theoretical efforts have been devoted in recent years to determine these quantities by fitting the pertinent experimental data~\cite{Landry:1999an,Landry:2002ix,DAlesio:2014mrz,Sun:2014dqm,Konychev:2005iy,Bacchetta:2017gcc,Scimemi:2017etj,Scimemi:2019cmh,Bacchetta:2019sam}, which, however, 
is limited by the imprecise knowledge of the nonperturbative behaviour of TMDPDFs and TMDWFs. Thus, it is highly desirable to develop a method to calculate them from first-principle approaches such as lattice QCD.

This has been realized in the framework of large momentum effective theory~\cite{Ji:2013dva,Ji:2014gla}, which offers a systematic way to calculate light-cone correlations by simulating time-independent Euclidean correlations on the lattice. Significant progress has been made in calculating various parton quantities from LaMET. For recent reviews, see Ref.~\cite{Cichy:2018mum,Ji:2020ect}. 

A very important result of LaMET development is that the TMDPDFs and TMDWFs can be calculated through the Euclidean quasi-TMDPDFs and quasi-TMDWFs, as well as a universal soft function (factor) ~\cite{Ebert:2019okf,Ji:2019sxk,Ji:2019ewn,Ji:2021znw}. In Ref.~\cite{Ji:2019sxk}, it has been suggested that the form factor of a bi-local four-quark operator calculable on the lattice, can be factorized into quasi-TMDWFs, a universal soft (function) factor and the matching kernel through QCD factorization at large momentum transfer, allowing
for the first time calculating the universal soft function on lattice. Thus the light-cone TMD parton distributions and wave functions can be obtained from numerical calculations of the four-quark form factors and quasi TMDPDFs and TMDWFs on the lattice~\cite{Ji:2019sxk,Ji:2019ewn}.  On the other hand, one can also make use of the QCD factorization to obtain the CS kernel from quasi TMDPDFs and TMDWFs. The first results for the CS kernel based on these proposals have been published recently~\cite{Shanahan:2020zxr,LatticeParton:2020uhz,Schlemmer:2021aij,Li:2021wvl,Shanahan:2021tst}. The quasi TMDWFs approach
to the CS kernel requires two-point function calculations 
and potentially can reach the light-cone limit with relatively
small hadron momenta. 

In this work, we present a state-of-the-art calculation of the CS kernel, based on a lattice QCD analysis of quasi-TMDWFs with $N_f=2+1+1$ valence clover fermions on a staggered quark sea with one-loop matching accuracy. A single ensemble with lattice spacing $a\simeq0.12$~fm,  volume $n_s^3\times n_t=48^3\times64$, and  physical sea-quark masses is used. In order to improve the signal-to-noise ratio, we tune the light-valence quark masses such that $m_\pi = 670$ MeV. The CS  kernel is then extracted through the ratios of the quasi-TMDWFs and the perturbative matching kernels at different momenta, $P^z=2\pi/n_s\times\{8,10,12\}=\{1.72,~2.15,~2.58\}$GeV. This  corresponds to Lorentz boost factors  $\gamma=\{2.57,~3.21,~3.85\}$, respectively. This analysis improves the previous ones~\cite{LatticeParton:2020uhz,Li:2021wvl} by taking into account  the one-loop perturbative contributions, and by analyzing  systematic uncertainties from operator mixing,  higher-order corrections from the scale dependence, and higher power corrections in terms of $1/P^z$.  
 
The remainder of this paper is organized as follows. 
In Sec.~\ref{sec:framework}, we present the theoretical framework to extract the CS kernel from  quasi-TMDWFs. Numerical results for quasi-TMDWFs and CS kernel are presented in Sec.~\ref{sec:numerics}. A brief  summary of this work is given in Sec.~\ref{sec:summary}.  More details about the analysis are collected in the appendix.


\section{Theoretical framework}
\label{sec:framework}
 
In this section, we review the necessary theoretical background
for the present calculation. We present the definitions of
CS kernel and rapidity evolution, and introduce the quasi-TMD wave functions. We then discuss the factorization of the quasi-TMDWFs
and its connection with the CS kernel. 
   
\subsection{Collins-Soper Kernel and Rapidity Evolution}

Unlike the collinear  lightcone  PDFs and  distribution amplitudes,  the TMDPDFs and TMDWFs depend on both the renormalization scale $\mu$ and an additional rapidity renormalization scale. The latter arises because the matrix elements also suffer from so-called rapidity divergences that require a dedicated regulator~\cite{Collins:1981uk,Becher:2010tm,Chiu:2011qc}.  In TMD factorizations, the contributions of hard, i.e. highly offshell, modes to the tree process are usually calculated in the dimensional regularization scheme. Collinear modes, which are related to  highly-boosted partons in distinct directions, and soft modes, whose typical momentum are  at the order  $\Lambda_{\mathrm{QCD}}$, share the same virtuality, 
and are only distinguishable by their rapidity. In calculations using regularization schemes such as dimensional regularization, which only regulate ultra-violet divergences, one will encounter additional rapidity divergences that arise in soft and collinear matrix elements when integrating over rapidity, and have to be resolved using a dedicated regulator. After the later regularization, TMDPDFs and TMDWFs acquire an additional rapidity scale dependence. This dependence should cancel in theoretical predictions for physical observables. 

The CS kernel $K\left(b_{\perp}, \mu\right)$, known as the rapidity anomalous dimension, encodes the rapidity dependence  of the TMD distributions~\cite{Collins:1981va,Collins:1981uk}:
\begin{align}
	2 \zeta \frac{d}{d \zeta} \ln f^{\mathrm{TMD}}\left(x, b_{\perp}, \mu, \zeta\right)=K\left(b_{\perp}, \mu\right), \label{eq:TMDsrapidityevolution}
\end{align}
where $f^{\mathrm{TMD}}$ denotes any leading twist TMDPDF or TMDWF. The TMD distributions depend on the longitudinal momentum fraction $x$, transverse spatial separation $b_{\perp}$, which is the Fourier-conjugate to the transverse momentum $k_{\perp}$, as well as the renormalization scale $\mu$ and rapidity scale $\zeta$ {which is related to the hadron momentum}. The $\mu$-dependence of CS kernel $K\left(b_{\perp}, \mu\right)$ satisfies the   renormalization group equation (RGE):
\begin{align}
	\mu^{2} \frac{d}{d \mu^{2}} K\left(b_{\perp}, \mu\right)=-\Gamma_{\text {cusp}}\left(\alpha_{s}\right).  \label{eq:cuspanomalousdimension}
\end{align}
Here $\Gamma_{\text {cusp}}\left(\alpha_{s}\right)=\alpha_sC_F/\pi+\mathcal{O}(\alpha_s^2)$ is the cusp anomalous dimension, which has been calculated  in  perturbation theory  up to 2-loop in Ref.~\cite{Li:2016ctv}, and 3-loop in Ref.~\cite{Moch:2017uml}. The solution to  the RGE  can be expressed as: 
\begin{align}
	K\left(b_{\perp}, \mu\right)=-2 \int_{1 / b_{\perp}}^{\mu} \frac{\mathrm{d} \mu^{\prime}}{\mu^{\prime}} \Gamma_{\text {cusp}}\left(\alpha_{s}\left(\mu^{\prime}\right)\right)+K\left(\alpha_{s}\left(1 / b_{\perp}\right)\right). 
	\label{eq:nonper-per-CS-kernel}
\end{align} 
For large $b_{\perp}$ with $b_{\perp}^{-1}\lesssim\Lambda_{\mathrm{QCD}}$, the CS kernel becomes nonperturbative, which is  represented by the non-cusp anomalous dimension $K\left(\alpha_{s}\left(1/b_{\perp}\right)\right)$ in the above equation. 

In the past decades, the CS kernel has  been widely studied in global fits of  TMD parton distributions~\cite{Landry:1999an,Landry:2002ix,DAlesio:2014mrz,Sun:2014dqm,Konychev:2005iy,Bacchetta:2017gcc,Scimemi:2017etj,Scimemi:2019cmh,Bacchetta:2019sam}. The explicit form in the nonperturbative region can only be parametrized by extending the perturbative expressions at  small $b_{\perp}$, which inevitably introduces systematic uncertainties. A direct calculation of TMDPDFs and the relevant CS kernel 
on the lattice was an almost insurmountable hurdle until the establishment  of LaMET~\cite{Ji:2013dva,Ji:2014gla}.  A remarkable recent development in LaMET is that  these quantities  can  be accessed through the corresponding quasi observables~\cite{Ebert:2019okf,Ji:2019sxk,Ji:2019ewn,Ji:2021znw}.


\subsection{Quasi TMD Wave Functions}
\begin{figure}
\centering
\includegraphics[scale=0.45]{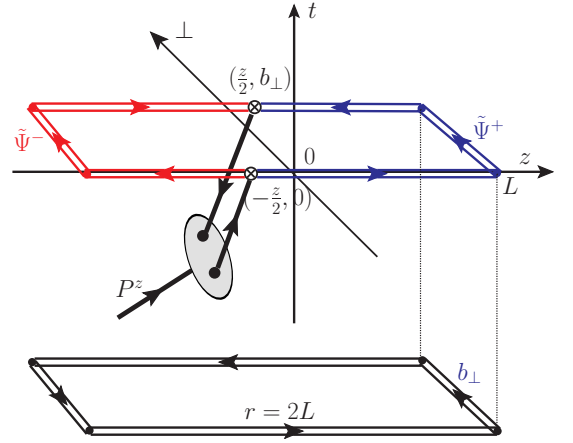}
\caption{Illustration of the staple-shaped gauge-link included in unsubtracted quasi-TMDWFs  and related Wilson loop. The blue and red double lines in the upper panel represent the $L$-shift direction on the Euclidean lattice, and the lower panel shows the correspondingly Wilson loop, which will subtract   UV logarithmic  and linear divergences in  quasi-TMDWFs . }
\label{fig:definitionofquasiTMDWF}
\end{figure}

As stated above, one can define the quasi TMDWFs for a highly-boosted pseudoscalar meson along the $z$-direction with large momentum $P^z$ as: 
\begin{align}
	\tilde{\Psi}^{\pm}&\left(x,b_{\perp},\mu,\zeta_z\right)=\lim_{L\to\infty}\int\frac{dz}{2\pi}e^{ix_rzP^z}\frac{\tilde{\Phi}^{\pm0}\left(z,b_{\perp},P^z,a,L\right)}{\sqrt{Z_E(2L,b_{\perp},\mu, a)}},\label{eq:quasiWFinmomentumspace}
\end{align}
where $x_r=x-\frac{1}{2}$. The unsubtracted quasi TMDWF $\tilde{\Phi}^{\pm0}$ is defined as an equal-time correlator containing nonlocal quark bilinear operator with staple-shaped gauge link:
\begin{align}
	\tilde{\Phi}^{\pm0}&\left(z,b_{\perp},P^z,a,L\right) =\left\langle 0\right|\bar{\psi}\left(z \hat{n}_{z}/2+b_{\perp} \hat{n}_{\perp}\right) \Gamma \nonumber\\
	&\times U_{\sqsupset, \pm L}\left(z \hat{n}_{z}/2+b_{\perp} \hat{n}_{\perp},-z \hat{n}_{z}/2\right) \psi\left(-z \hat{n}_{z}/2\right)\left| P^{z}\right\rangle. \label{eq:quasiWFincoordinatespace}
\end{align}
For a pseudoscalar mesonic state, the Dirac structure $\Gamma$ can be chosen as $\gamma^z\gamma_5$ or $\gamma^t\gamma_5$, which approaches the leading-twist structure $\gamma^+\gamma_5$ in the light-cone limit. With a large but finite $P^z$, the difference between $\gamma^z\gamma_5$ and $\gamma^t\gamma_5$  is  suppressed by powers of $1/P^z$. Technically, one can also use a combination of them, such as $\left(\gamma^z\gamma_5+\gamma^t\gamma_5\right)/2$ to minimize  power corrections, and more details can be found  in Sec.~\ref{sec:operator}.  Various combinations were also explored in Ref.~\cite{Li:2021wvl}.  The superscript ''0'' in $\tilde{\Phi}^{\pm0}$ indicates bare quantities. The linear divergences come from the self-energy of the gauge-link, 
\begin{align}
	&U_{\sqsupset, \pm L}\left(z \hat{n}_{z}/2+b_{\perp} \hat{n}_{\perp},-z \hat{n}_{z}/2\right) \equiv\nonumber\\
	&	U_z^{\dagger}\left(z \hat{n}_{z}/2+b_{\perp} \hat{n}_{\perp}; L \right) U_{\perp}\big((L-z /2)\hat{n}_{z};{b}_T \big) U_z\left( -z \hat{n}_{z}/2; L \right), \label{eq:stapleshapedUlink}
\end{align}
and does not appear as a pole at $d=4$ in dimensional regularization. The Euclidean gauge link in $U_{\sqsupset, \pm L}$ is defined as
\begin{align}
	U_z(\xi,\pm L)=\mathcal{P}	\exp \left[-i g \int_{\xi^{z}}^{\pm L} d \lambda n_{z} \cdot A\left(\vec{\xi}_{\perp}+n_{z} \lambda\right)\right],
\end{align}
where $\xi^{z}=-\xi \cdot n_{z}$. The $\pm L$  corresponds to the farthest position that the gauge link can reach in positive or negative $n_z$ direction  on a finite Euclidean lattice. This is depicted as the blue and red lines in Fig.~\ref{fig:definitionofquasiTMDWF}.

Since the  linear divergence is associated with the gauge-link, it can be removed by a similar gauge-link with the same total length. An  optional choice  is to make use the Wilson loop,  denoted as $Z_E$. The Wilson loop can be chosen  as the vacuum expectation of a flat rectangular Euclidean Wilson-loop in the $z$-$\perp$ plane:
\begin{align}\label{eq:WilsonLoop}
Z_E\left(2L, b_{\perp}, \mu,a\right)=\frac{1}{N_{c}} \operatorname{Tr}\left\langle 0\left|U_{\perp}(0;b_{\perp}) U_{z}\left(b_{\perp}\hat{n}_{\perp}; 2L\right)\right| 0\right\rangle. 
\end{align}
Here  the length of $Z_E$ is twice that of the staple-shaped gauge-link $U_z^{(\dagger)}$ in the $z$ direction, and thus it is anticipated that the square root of $Z_E\left(2L, b_{\perp}, \mu\right)$ cancels the linear divergence and heavy quark potential in the gauge-link.  There are residual logarithmic divergences from the vertex of   Wilson line and light quark, which can be renormalized in dimensional regularization~\cite{Ji:2021uvr}. As these logarithmic divergences are independent of $z$, $b_{\perp}$, $P_{z}$ and $L$, they will explicitly cancel out when  the ratio of quasi TMDWFs is studied.

\subsection{Factorization of Quasi TMDWFs}\label{sec:factorizationformula}

With the help of soft function, the infrared contributions in the subtracted quasi TMDWFs can be properly accounted for such that the infrared structures for the   quasi TMDWFs and light-cone ones are matched. This implies a multiplicative factorization theorem in the framework of LaMET~\cite{Ebert:2019okf,Ji:2019sxk,Ji:2019ewn,Ji:2021znw,Ebert:2022fmh}:
\begin{align}
&\tilde \Psi^{\pm}(x, b_{\perp}, \mu, \zeta_z) S_r^{1/2}(b_{\perp}, \mu) \nonumber\\
&= H^{\pm}\left(\zeta_z, \overline\zeta_z, \mu^2\right) \exp\left[\frac{1}{2}K(b_{\perp},\mu)\ln\frac{\mp\zeta_z- i\epsilon}{\zeta} \right]  \nonumber\\
& \qquad\times \Psi^{\pm}(x, b_\perp, \mu, \zeta)+\mathcal{O}\left(\frac{\Lambda_{QCD}^2}{\zeta_z}, M^2\zeta_z,\frac{1}{b_{\perp}^2\zeta_z}\right),
\label{eq:matching}
\end{align}
where the superscript $\pm$ in Eq.~\eqref{eq:matching} corresponds to the direction in the Wilson line, $\Psi^{\pm}$ is the TMDWFs defined in the inﬁnite momentum frame. The reduced soft function $S^{1/2}_r(b_{\perp}, \mu)$ emerges from the different soft gluon radiation effects in $\tilde \Psi^{\pm}$ and $\Psi^{\pm}$~\cite{Ji:2021znw}. The mismatch  of the rapidity scale $\zeta$ and $\zeta_z$ can be compensated by the CS kernel $K(b_{\perp},\mu)$. Both $S$ and $K$ are independent of the $\pm$ choice.  $H^{\pm}$ is the 1-loop perturbative matching kernel~\cite{Ji:2021znw}:
\begin{align}
& H^{\pm}(\zeta_z, \overline{\zeta}_z,\mu) \nonumber\\
& = 1+  \frac{ \alpha_s C_F}{4\pi} \bigg(-\frac{5\pi^2}{6}-4 +{\ell}_{\pm} + \overline \ell_{\pm} - \frac{1}{2} ({\ell}_{\pm}^2 + \overline \ell_{\pm}^2) \bigg).  \label{eq:1loophardkernel}
\end{align}
With the abbreviations ${\ell}_{\pm} = \ln \left[(-\zeta_z \pm i\epsilon)/\mu^2\right]$, and $\overline {\ell}_{\pm} = \ln \left[(-\overline \zeta_z \pm i\epsilon)/\mu^2\right]$, the scales $\zeta_z = (2x P^z)^2$ and $\overline\zeta_z = \left(2\bar{x} P^z\right)^2$, and $\bar{x}=1-x$. It should be noticed that $H^{\pm}$ contains nonzero imaginary parts in $\ell_{\pm}$ and $\bar{\ell}_{\pm}$. While the imaginary parts in  $\ell_{\pm}/\bar{\ell}_{\pm}$ are constants, the  ones in the double logarithms ${\ell}_{\pm}^2/\bar{\ell}_{\pm}^2$ are momentum-dependent.

\begin{figure}
\centering
\includegraphics[scale=0.4]{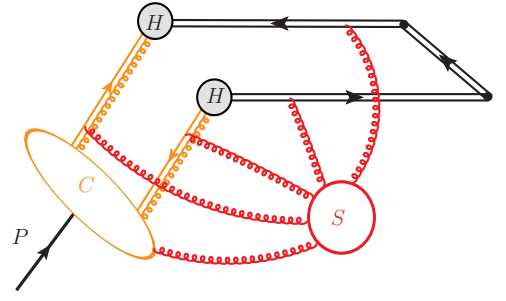}
\caption{Leading-power reduced graph for pseudoscalar meson quasi TMDWFs. Here $C, ~S$ denote the collinear and soft sectors of the infra-red structure, while $H$ denote the hard contributions. Since the hard-gluon exchange between the quark and anti-quark is power suppressed, the hard parts are disconnected with each others, and therefore the factorization of quasi-WF amplitude is multiplicative.  }
\label{fig:reduced_graph}
\end{figure}

A characteristic behavior of Eq.(\ref{eq:matching}) is that this factorization is multiplicative~\cite{Ji:2020ect}, which indicates that hard gluon contributions are local. This is due to the fact that hard gluon exchange between  the quark and anti-quark  sectors in quasi TMDWFs is power suppressed: if there were such a hard gluon, the spatial separation between its attachments  is much smaller than $b_{\perp}$, resulting in power suppression compared to the typical hard mode contributions.  
Thus, at leading power the factorization of quasi TMDWFs are multiplicative. This feature  is illustrated in Fig.\ref{fig:reduced_graph}, in which the collinear, soft and hard sub-diagrams represent the pertinent  contributions. Further, $\zeta_z$ arising from   Lorentz-invariant combinations of collinear momentum modes, will provide the natural hard scale of the hard sub-diagram. More detailed explanations for the factorization of quasi TMDPDFs in LaMET are given in the recent review~ \cite{Ji:2019ewn}.

\subsection{Collins-Soper Kernel From Quasi TMDWFs}\label{sec:extractingCSkernel}

From Eq.(\ref{eq:matching}), one can see that the momentum dependence in   quasi TMDWFs provides an option to determine the CS kernel. This can be written  in a way similar to Eq.(\ref{eq:TMDsrapidityevolution})~\cite{Ji:2021znw}: 
\begin{align}
	2 \zeta_z \frac{d}{d \zeta_z} \ln \tilde \Psi^{\pm}\left(x, b_{\perp}, \mu, \zeta_z\right)=&K\left(b_{\perp}, \mu\right) +\frac{1}{2}\mathcal{G}^{\pm}\left(x^2\zeta_z,\mu\right) \nonumber\\
	+& \frac{1}{2}\mathcal{G}^{\pm}\left(\bar{x}^2\zeta_z,\mu\right)+\mathcal{O}\left(\frac{1}{\zeta_z}\right), \label{eq:quasiTMDsrapidityevolution}
\end{align}
where $K\left(b_{\perp}, \mu\right)$ denotes the same kernel as in Eq.(\ref{eq:TMDsrapidityevolution}), and does not depend on the hard scale $\zeta_z$ for large $P^z$. Unlike TMDWFs, the  quasi distributions also contain hard contributions, whose  rapidity dependence is  represented by the perturbative  $\mathcal{G}^{\pm}$ as a function of hard scale $\zeta_z$. 
From the $\zeta_z$ dependence of quasi TMDWFs, we can see that when $P^z\to\infty$, the large logarithms in $P^z$ are partially absorbed into $K\left(b_{\perp}, \mu\right)$  and the remanent is incorporated in the perturbative matching kernel. Therefore,  both the matching kernel $H^{\pm}$  and an exponential of  CS kernel $K\left(b_{\perp}, \mu\right)$ are needed to describe the dependence on $\zeta_z$ of quasi TMDWFs. 

In order to extract the CS kernel $K\left(b_{\perp}, \mu\right)$ explicitly, one  can make use of    Eq.(\ref{eq:matching}) with  two different large momenta $P_1^z\neq P_2^z\gg1/b_{\perp}$ but the same   scale $\zeta_z$. Taking a  ratio of these two quantities  gives 
\begin{align}
	\frac{\tilde \Psi^{\pm}(x, b_{\perp}, \mu,P_1^z)}{\tilde \Psi^{\pm}(x, b_{\perp}, \mu, P_2^z)}=\frac{H^{\pm}\left(xP_1^z,\mu \right)}{H^{\pm}\left(xP_2^z,\mu \right)}\exp\left[K(b_{\perp},\mu) \ln\frac{P_1^z}{P_2^z}\right], \label{eq:ratioofquasiWF}
\end{align}
where the reduced soft function $S_r(b_{\perp},\mu)$ and TMDWFs $\Psi^{\pm}(x,b_{\perp},\mu,\zeta)$  have been canceled in the ratio. Therefore, the CS kernel $K\left(b_{\perp}, \mu\right)$ can be extracted through 
\begin{align}
	K\left(b_{\perp}, \mu\right)=\frac{1}{\ln(P_1^z/P_2^z)} \ln\frac{H^{\pm}(xP_2^z,\mu)\tilde{\Psi}^{\pm}(x,b_{\perp},\mu,P_1^z)}{H^{\pm}(xP_1^z,\mu)\tilde{\Psi}^{\pm}(x,b_{\perp},\mu,P_2^z)}. \label{eq:extractingCSkernel}
\end{align}
Note that the extracted result is formally independent of $x$ and $P_{1/2}^z$ at leading power,  and both $\tilde{\Psi}^{+}$ and $\tilde{\Psi}^{-}$ can be used to extract $K\left(b_{\perp}, \mu\right)$.  This is derived at the leading power in the factorization scheme and  might be undermined by power corrections. Accordingly, in order  to reduce  the systematic uncertainties,  we take the average: 
 \begin{align}
	K\left(b_{\perp}, \mu\right)=&\frac{1}{2\ln(P_1^z/P_2^z)} \left[\ln\frac{H^{+}(xP_2^z,\mu)\tilde{\Psi}^{+}(x,b_{\perp},\mu,P_1^z)}{H^{+}(xP_1^z,\mu)\tilde{\Psi}^{+}(x,b_{\perp},\mu,P_2^z)}\right.  \nonumber\\
	&+\left.\ln\frac{H^{-}(xP_2^z,\mu)\tilde{\Psi}^{-}(x,b_{\perp},\mu,P_1^z)}{H^{-}(xP_1^z,\mu)\tilde{\Psi}^{-}(x,b_{\perp},\mu,P_2^z)}\right].
\end{align}
The details will be discussed in Sec. \ref{subsec:csresults}.

\section{Numerical Simulations and Results}
\label{sec:numerics}

In this section, we present our lattice QCD results. We start with the lattice setup, followed by results
for quasi TMDWFs with two-point correlations. 
The Wilson loop results are discussed in subsection III.C. Subsection III.D studies the operator mixing effects. Our main result on CS kernel is presented in subsection E. The final subsection E includes some overall discussions. 

\subsection{Lattice setup }

Our numerical simulations use $N_f=2+1+1$ valence clover fermions on a  highly improved staggered quark (HISQ) sea~\cite{Follana:2006rc} and a 1-loop Symanzik improved gauge action~\cite{Symanzik:1983dc}, generated by the MILC collaboration~\cite{MILC:2012znn} using periodic boundary conditions. In the calculations, we use a single ensemble with the lattice spacing $a\simeq0.12$~fm and the volume $n_s^3\times n_t=48^3\times64$ at physical sea-quark masses. In order to increase the signal-to-noise ratio, we tune the light-valence quark masses to the strange-quark one, namely $m_{\pi}^{\mathrm{sea}}=130$~MeV and  $m_{\pi}^{\mathrm{val}}=670$~MeV, which
could generate some non-unitarity effects. On the other hand, the Collins-Soper kernel only depends weakly on quark mass, and we may consider valence quarks are strange-like, namely the hadrons involved are kaons.  

To further improve the statistical signals, we adopt hypercubic (HYP) smeared fat links~\cite{Hasenfratz:2001hp} for the gauge ensembles. To access the large momentum limit for the CS kernel, we employ three different hadron momenta as $P^z=2\pi/n_s\times\{8,10,12\}=\{1.72,~2.15,~2.58\}$~GeV  corresponding  to the boost factor $\gamma=\{2.57,~3.21,~3.85\}$. 


\subsection{Quasi TMDWFs From Two-Point Correlators}
\label{sec:quasi-WF_from_2pt}
In order to calculate the quasi TMDWFs  defined in Eq.(\ref{eq:quasiWFinmomentumspace}), we generate Coulomb-gauge wall-source propagators, 
\begin{align}
S_{w}\left(x, t, t^{\prime} ; \vec{p}\right)=\sum_{\vec{y}} S\left(t, \vec{x} ; t^{\prime}, \vec{y}\right) e^{i \vec{p} \cdot(\vec{y}-\vec{x})}, \label{eq:wallsource}
\end{align}
where $(t^{\prime},\vec{y})$ and $(t,\vec{x})$ denote the space-time positions of source and sink. Then one  can construct the two-point function (2pt) related to the quasi TMDWFs  in Eq.~(\ref{eq:quasiWFinmomentumspace}):
\begin{align}
C_2^{\pm}&	\left(z,b_{\perp},P^z; p^z,L,t \right)=\frac{1}{n_s^3}\sum_{\vec{x}}\mathrm{tr} e^{i\vec{P}\cdot\vec{x}}\left\langle S_w^{\dagger}\left(\vec{x}_1,t,0;-\vec{p}\right)\right. \nonumber\\
&\;\;\;\;\;\;\;\;\;\; \;\;\;\;\;\;\;\;\;\;    \times \left. \Gamma U_{\sqsupset, \pm L}\left(\vec{x}_1,\vec{x}_2\right) S_w\left(\vec{x}_2,t,0;\vec{p} \right)  \right\rangle
\label{eq:nonlocal2pt}
\end{align}
with $\vec{x}_1=\vec{x}+z\hat{n}_z/2+b_{\perp}\hat{n}_{\perp}$ and  $\vec{x}_2=\vec{x}-z \hat{n}_{z}/2$. The quark momentum $\vec{p}=(\vec{0}_{\perp},p^z)$ is along the $z$-direction, and each of two quarks carries half of the hadron momentum. Thereby the hadron momentum satisfies  $\vec{P}=2\vec{p}$. The anti-quark propagator can be obtained from Eq.(\ref{eq:1loophardkernel}) by applying $\gamma_5$-hermiticity $S_w(x,y)=\gamma_5S_w(y,x)^{\dagger}\gamma_5$. As mentioned above, the Dirac structures are chosen as $\Gamma=\gamma^z\gamma_5$ and $\gamma^t\gamma_5$ that can be projected onto the leading twist light-cone contributions in the large $P^z$ limit.
 
By generating the wall-source propagators with quark momenta $p^z=\pm\{4,5,6\}\times2\pi/(n_sa)$, and three segments of gauge links following Eq.(\ref{eq:stapleshapedUlink}), one  can construct  the two-point correlation functions on the  lattice. With the help of reduction formulas, the $C_2^{\pm}$ can be parametrized as
 \begin{align}
C_2^{\pm}&\left(z,b_{\perp},P^z; p^z,L,t \right)=\frac{A_w(p^z)A_p}{2E}\tilde{\Phi}^{\pm0}\left(z,b_{\perp},P^z,L\right) \nonumber\\
& \times e^{-Et}\left[1+c_0\left(z,b_{\perp},P^z,L\right)e^{-\Delta Et} \right], \label{eq:C2parametrization}
 \end{align}
where $A_w(p^z)$ is the matrix element of the pseudoscalar meson interpolating ﬁeld with Coulomb gauge fixed wall source and $A_p$ is the one for a point source (sink). These terms as well as the factor $E^{-1}=1/\sqrt{m_{\pi}^2+\left(P^z\right)^2}$ are cancelled by the local  two-point function $C_2^{\pm}\left(0,0,P^z; p^z,0,t \right)$ at the same time slice. Thus the remaining  ground-state matrix element $\tilde{\Phi}^{\pm0}$ is normalized. The ratio of nonlocal and local two-point functions can be parametrized  as
\begin{align}
&R^{\pm}\left(z,b_{\perp},P^z,L,t \right)=\frac{C_2^{\pm}\left(z,b_{\perp},P^z,L,t \right)}{C_2\left(0,0,P^z,0,t \right)} \nonumber\\
=&\tilde{\Phi}^{\pm0}\left(z,b_{\perp},P^z,L\right)\left[1+c_0\left(z,b_{\perp},P^z,L\right)e^{-\Delta Et} \right].
\label{eq:two-state-fit}
\end{align}
where $C_2$ in the denominator is a local
correlator.
 
\begin{figure}
\centering
\includegraphics[scale=0.55]{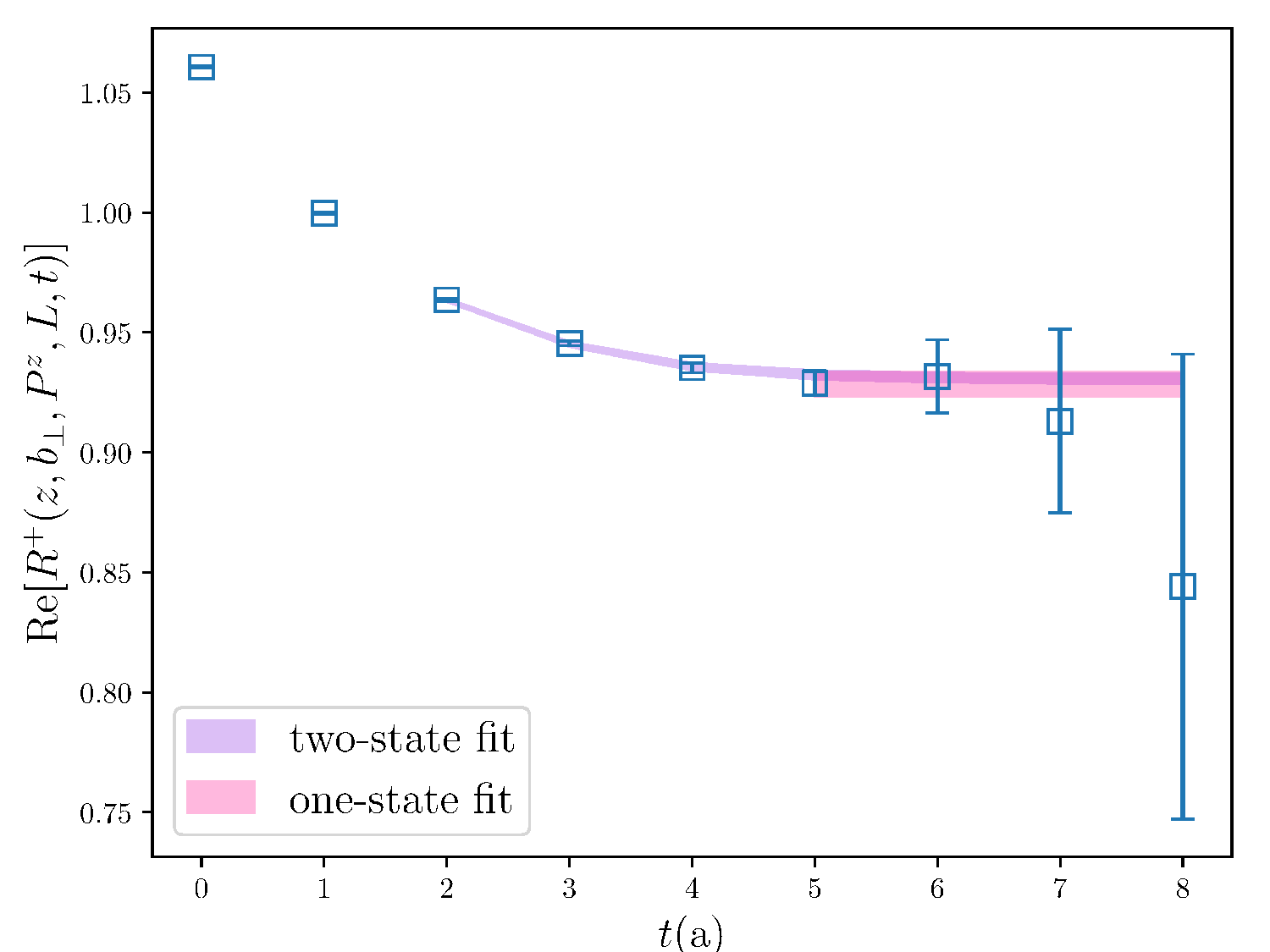}
\caption{Comparison of  two-state fit  and  one-state fit to extract $\tilde{\Phi}^{\pm0}\left(z,b_{\perp},P^z,L\right)$. Taking $\{z,b_{\perp},P^z,L\}=\{0a, 2a, 24\pi/n_s, 6a\}$ as example, we can see that  the two-state fit works for $t\in[2a, 8a]$ while the one-state fit works for $t\in[5a, 8a]$. The fitted  results are consistent with each other, while  the one-state fit is more conservative. }
\label{fig:tdependenceofratio}
\end{figure}

In the above parametrization,  the excited-state contributions are collected into the $c_0$ term, and $\Delta E$ denotes the mass gap between the ground and first excited state.  With the increase of Euclidean time, contributions from the excited state decay and the plateau obtained for $R^{\pm}\left(z,b_{\perp},P^z,L,t \right)$ at large times reflects the ground-state contribution $\tilde{\Phi}^{\pm0}$. We employ two methods to extract $\tilde{\Phi}^{\pm0}$, namely the two-state fit directly using Eq.~\eqref{eq:two-state-fit}, and the one-state fit by setting $c_0=0$. With a large enough Euclidean time, Fig.~\ref{fig:tdependenceofratio} exhibits a comparison using two methods for the case with small $\{z,b_{\perp}\}$. From this figure, one can see that the one-state fit result is consistent with two-state fit one but gives  a more conservative error estimate. The two-state fit works at $t\in[2a, 8a]$ and the one-state fit works at the plateau region $t\in[5a, 8a]$. While the excited state contamination would dominate for the two-state fit in a very high precision especially for the cases with large $\{z, b_\perp\}$. So with current accuracy, we adopt the more conservative results from the one-state fit in the following analysis. More details can be found in Appendix~\ref{appsec:C2}.

\subsection{Wilson Loop Renormalization}
\label{subsec:numerical_wl}

The unsubtracted quasi TMDWF matrix elements $\tilde{\Phi}^{\pm0}\left(z,b_{\perp},P^z,a,L\right)$ extracted from the joint fit of the  two-point function contain a factor $e^{-\delta \bar{m}(2L+b_{\perp})}$ from the linear divergence, the heavy quark effective potential factor $e^{-V(b_{\perp})L}$ and logarithmic divergences $Z_{O}$:
\begin{align}
\tilde{\Phi}^{\pm0}\left(z,b_{\perp},P^z,a,L\right) \propto e^{-\delta \bar{m}(2L+b_{\perp})}e^{-V(b_{\perp})L}Z_{O}.
\end{align} 
where $Z_O$ has logarithmic dependence on lattice spacing $a$. 

The linear divergence in $e^{-\delta \bar{m}(2L+b_{\perp})}$ comes from the self-energy of the Wilson line~\cite{Ji:2017oey,Ishikawa:2017faj,Green:2017xeu, Ji:2020brr}, where $\delta\bar m$ contains a term  proportional to $1/a$  and a  non-perturbative renormalon contribution $m_{0}$:
\begin{align}
\delta\bar m = \frac{m_{-1}(a)}{a} - m_0  \ . 
\end{align}
Note that the exponent of the linear divergence term is proportional to the total length of the Wilson link, e.g. $2L+b_{\perp}$ for the staple link. Due to this factor,  the numerical value for a Wilson loop dramatically decreases  for small $a$ and large $L$. 

The heavy quark effective potential term $e^{-V(b_{\perp})L}$ comes from interactions between the two Wilson lines along the $z$ direction in the staple link. The heavy quark effective potential $V(b_{\perp})$ is often used to determine the lattice spacing of an ensemble.  

The logarithmic divergence $Z_{O}$ comes from the vertices
involving the Wilson line and light quark. The logarithmic  divergence up to leading order, resumed by renormalization group equation, and matched to  $\overline{\rm MS}$ scheme is~\cite{Ji:1991pr, LatticePartonCollaborationLPC:2021xdx} 
\begin{align}\label{eq:ZO_RS_higher}
Z_{O}(1/a, \mu)=\left(\frac{\ln [1 /(a \Lambda_{\rm QCD}^{\rm latt})]}{\ln [\mu / \Lambda_{\rm QCD}^{\rm MS}]}\right)^{\frac{3 C_{F}}{b_0}},\nonumber\\
\end{align}
where $\Lambda_{\rm QCD}^{\rm latt}$  is different from that in $\overline{\rm MS}$ scheme. One can use both to effectively absorb higher order contributions~\cite{Lepage:1992xa}.

\begin{figure}
\centering
\includegraphics[scale=0.55]{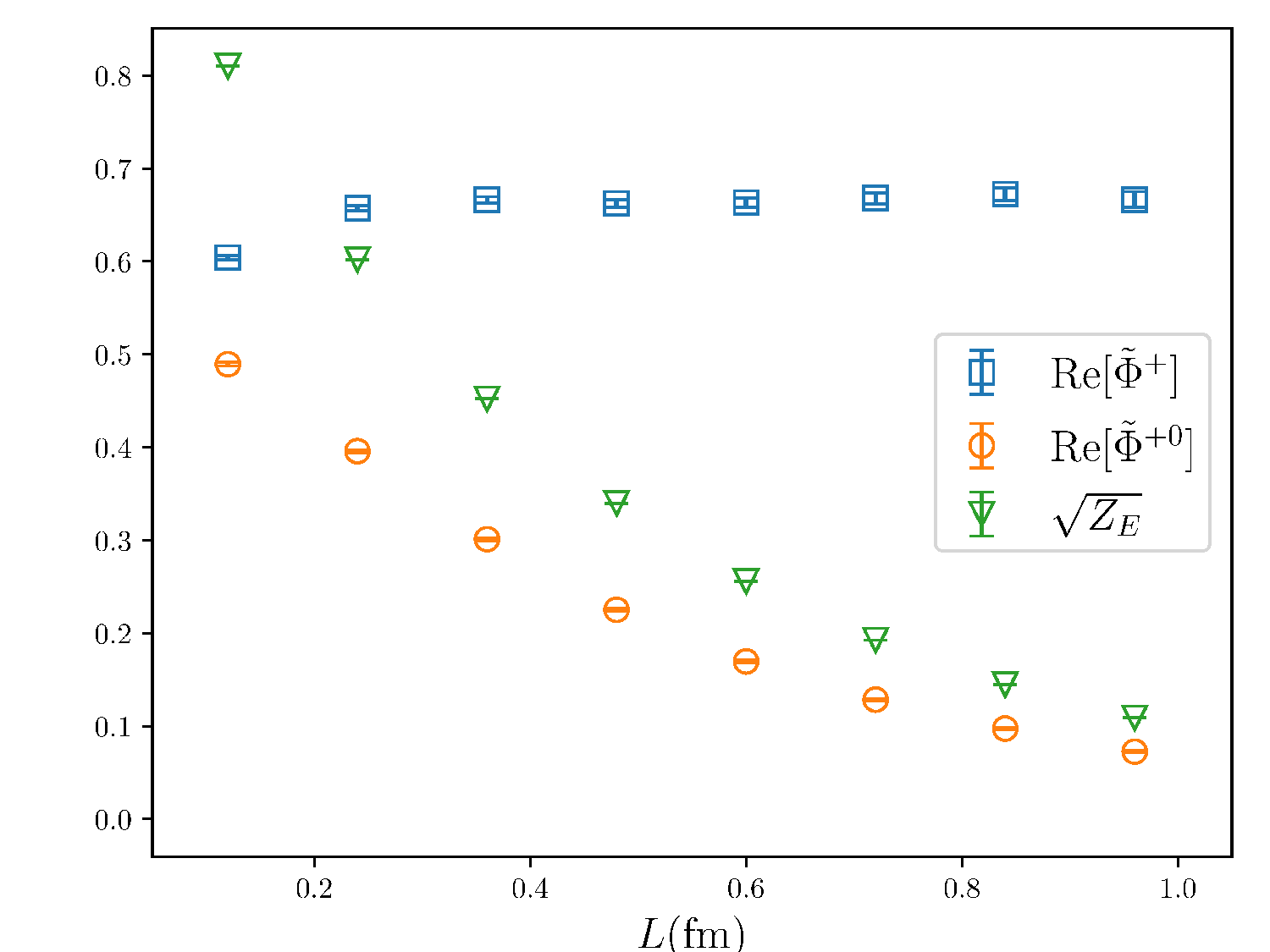}
\includegraphics[scale=0.55]{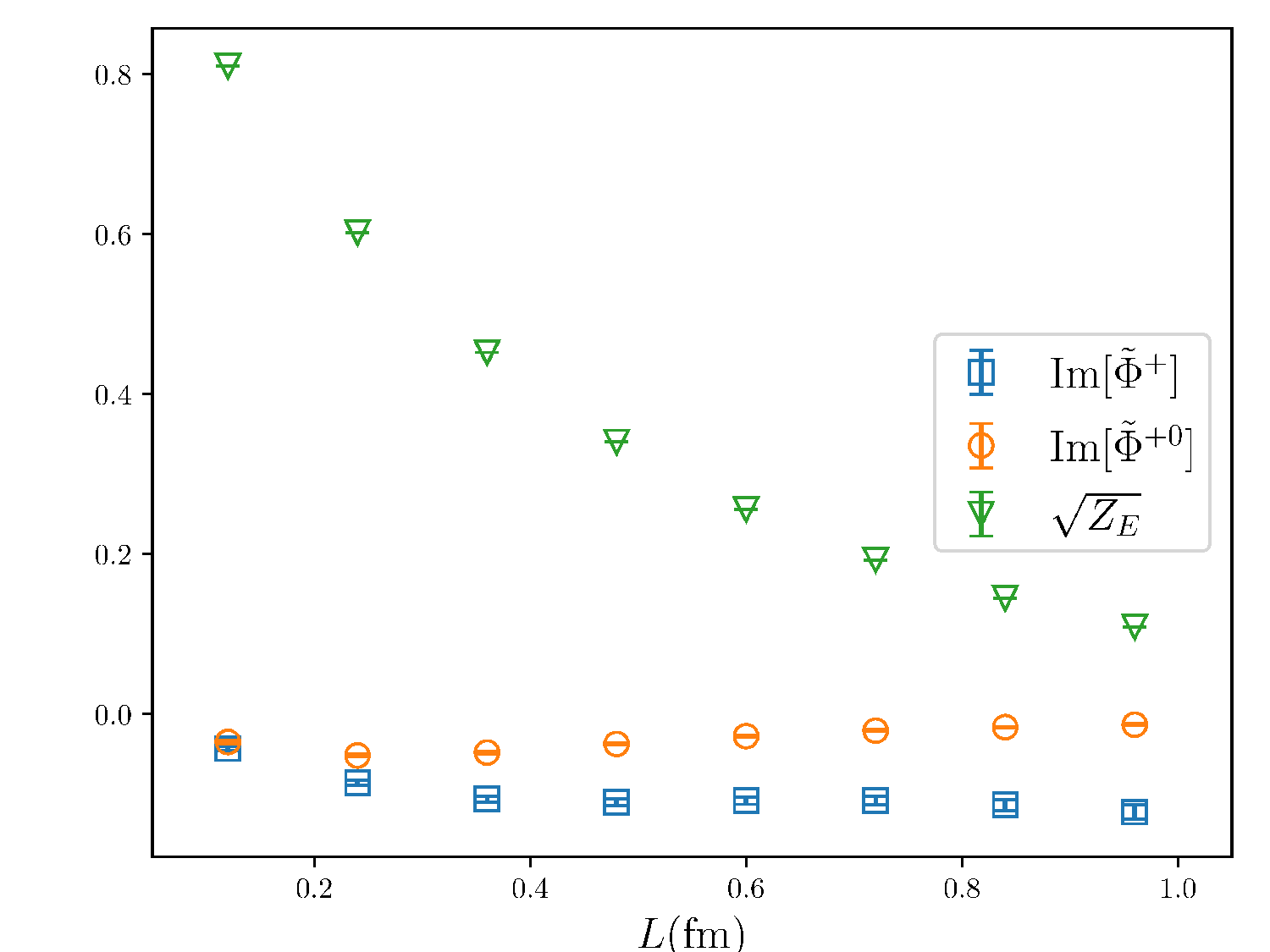}
\caption{Results for the $L$-dependence of  unsubtracted and subtracted quasi TMDWFs:  real part (upper panel) and imaginary part (lower panel), as well as the square root of the Wilson loop. The case $\Gamma=\gamma^z\gamma_5$ and $\{P^z,b_{\perp}, z\}=\{16\pi/n_s,2a,2a\}$ is used for illustration. }
\label{fig:WilsonLooprenormalization}
\end{figure}

In this work,  the Wilson loop renormalization method~\cite{Chen:2016fxx,Zhang:2017bzy,Musch:2010ka,Green:2017xeu,Zhang:2017zfe}  is adopted, in which the Wilson loop $Z_E$ defined in Eq.(\ref{eq:WilsonLoop})  contains linear divergence and heavy quark effective potential:
\begin{align}
Z_E\left(2L, b_{\perp}, a\right) \propto e^{-\delta \bar{m}(4L+2b_{\perp})}e^{-V(b_{\perp})2L}.
\end{align}
According to Ref.~\cite{LatticePartonCollaborationLPC:2021xdx}, $\delta \bar{m}$ in the Wilson loop is the same as that in hadron matrix elements, and thus it is anticipated that the linear divergence is removed when dividing by $\sqrt{Z_E\left(2L, b_{\perp}, a\right)}$:
\begin{align}
\tilde{\Phi}^{\pm}(z,b_{\perp},P^z,a,L)=\frac{\tilde{\Phi}^{\pm0}\left(z,b_{\perp},P^z,a,L\right)}{\sqrt{Z_E\left(2L, b_{\perp}, a\right)}}.
\end{align}
As shown in Fig.~\ref{fig:WilsonLooprenormalization}, the subtracted quasi TMDWFs tend  to  be a constant when  $L\ge 0.4$ fm. We then use the subtracted quasi TMDWFs defined as
\begin{align}\label{eq:WLRQuasi}
\tilde{\Phi}^{\pm}(z,b_{\perp},P^z, a)=\lim_{L\to\infty}\frac{\tilde{\Phi}^{\pm0}\left(z,b_{\perp},P^z,a,L\right)}{\sqrt{Z_E\left(2L, b_{\perp}, a\right)}}.
\end{align}
However, it is anticipated that there is a residual logarithmic divergence $Z_{O}$:
\begin{align}
\frac{\tilde{\Phi}^{\pm0}\left(z,b_{\perp},P^z,a,L\right)}{\sqrt{Z_E\left(2L, b_{\perp}, a\right)}}
\propto \frac{e^{-\delta \bar{m}(2L+b_{\perp})}e^{-V(b_{\perp})L}Z_{O}}{e^{-\delta \bar{m}(2L+b_{\perp})}e^{-V(b_{\perp})L}}
= Z_{O}.
\end{align}
In the extraction of the CS kernel, a ratio of quasi TMDWFs is adopted and accordingly the residual logarithmic divergence $Z_{O}$ is canceled.

\subsection{Higher-Twist Effects in Operators}
\label{sec:operator}

\begin{figure}
\centering
\includegraphics[scale=0.55]{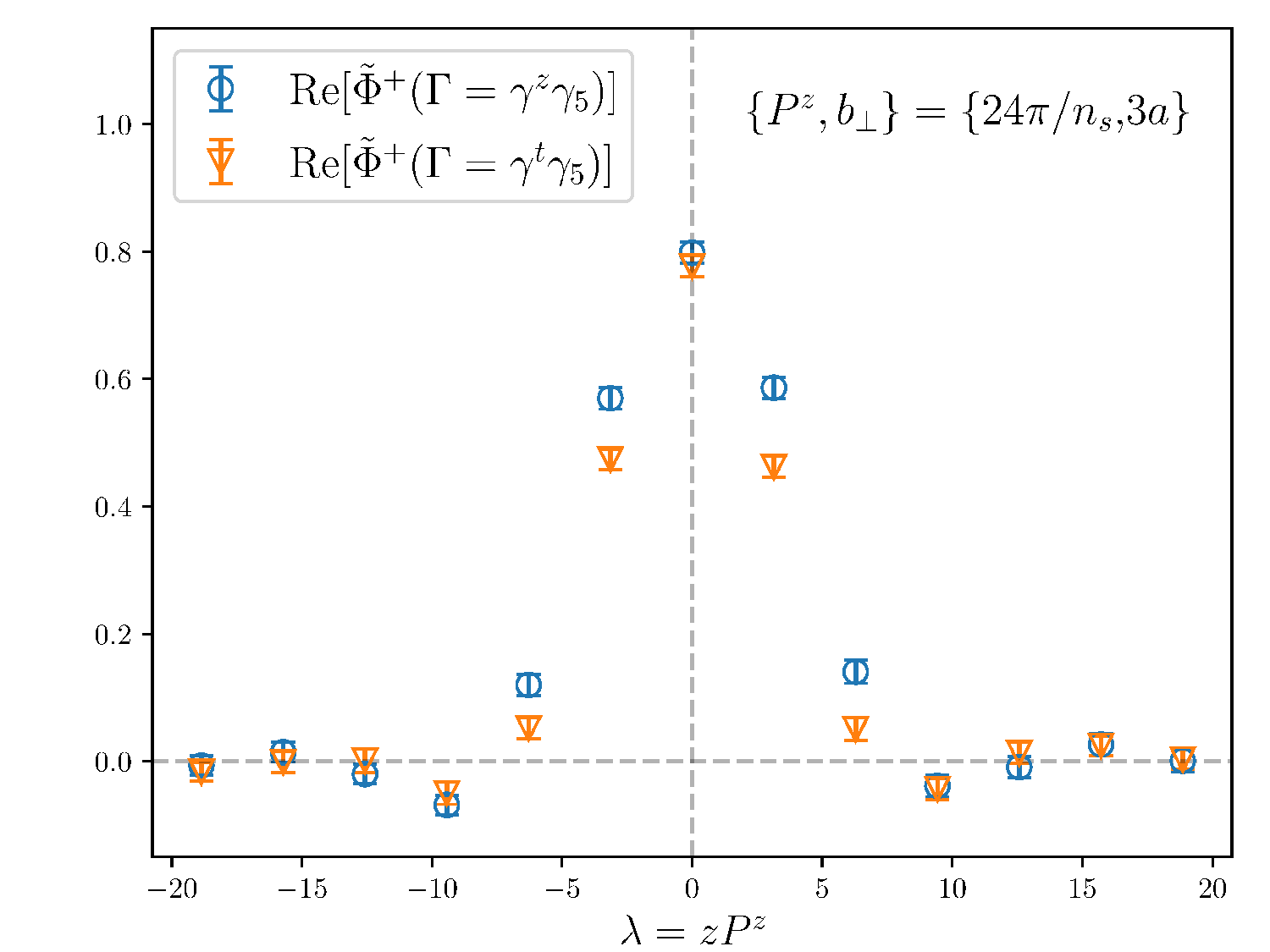}
\includegraphics[scale=0.55]{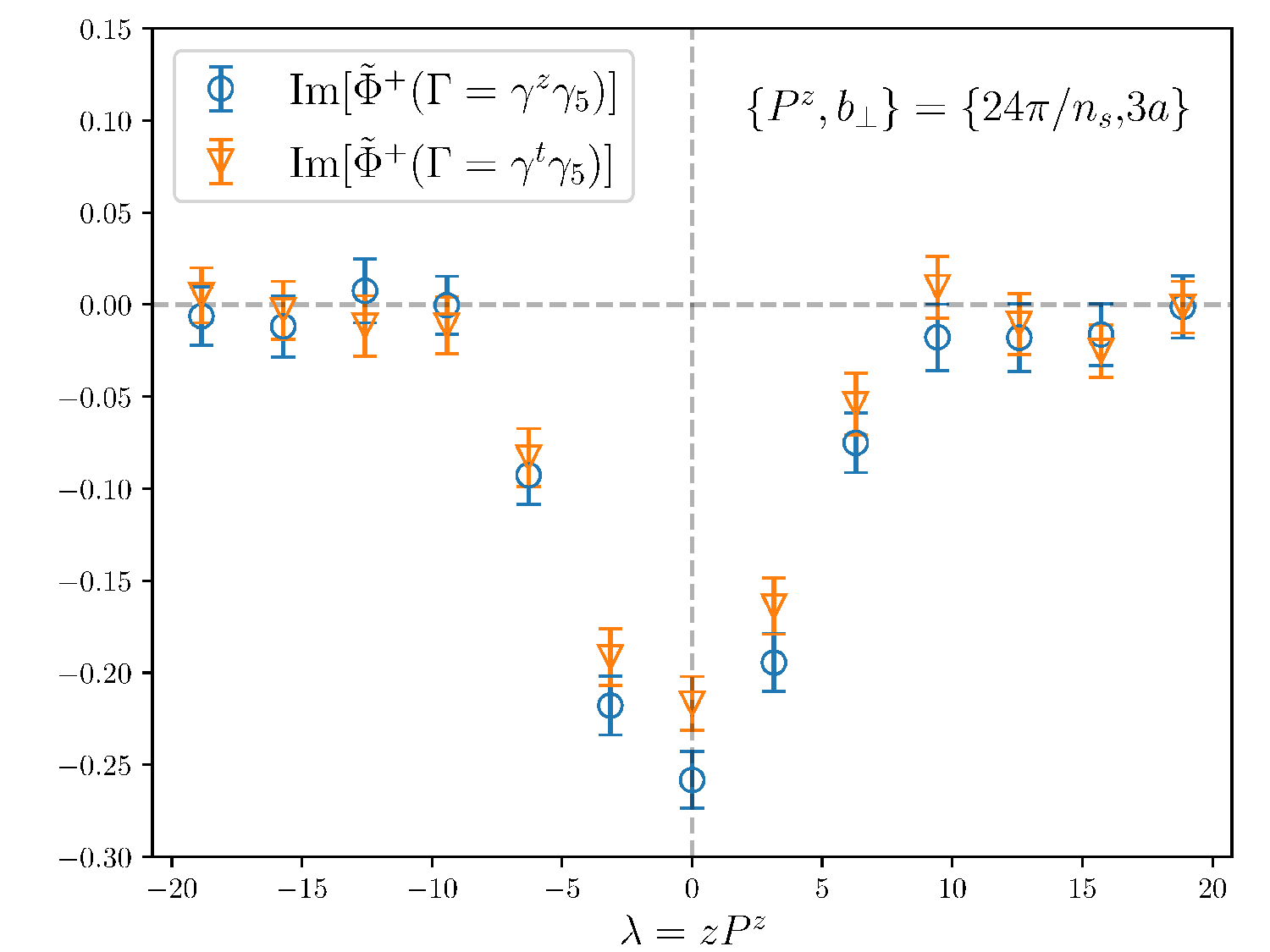}
\caption{$\lambda$-dependence of quasi-WF matrix elements with different Dirac structures. Here we take the case of $\{P^z,b_{\perp}\}=\{24\pi/n_s,3a\}$ as an example, the deviation between these two cases mainly comes from power corrections. Results for more sets of $\{P^z,b_{\perp}\}$ are shown in appendix~\ref{sec:app_operator_mixing}.}
    \label{fig:qwf_coordinate}
\end{figure}

For a pseudoscalar meson on a Euclidean lattice, both $\Gamma=\gamma^t\gamma_5$ and $\gamma^z\gamma_5$ project onto the leading twist light-cone distribution amplitude, i.e. $\gamma^+\gamma_5$ in the arge $P^z$ limit. The differences  between them arises from power corrections in terms of $M^2/\left(P^z\right)^2$.

Fig.~\ref{fig:qwf_coordinate} shows the comparison of the $\lambda= z   P^z$-dependence of quasi TMDWFs with $\Gamma=\gamma^t\gamma_5$ and $\gamma^z\gamma_5$ at $P^z=24\pi/n_s\simeq2.58$~GeV. It can be seen from the plots that there are some differences between the two sets of results for the real part in  the small $\lambda$ region. The differences are expected to decrease with increasing $P^z$, and the correlators with $\gamma^t\gamma_5$ and $\gamma^z\gamma_5$ will gradually converge to the light-cone from opposite directions. Besides, in light-cone coordinate, $\gamma^t$ and $\gamma^z$ can be represented by $\gamma^+$ and $\gamma^-$,
\begin{align}
\gamma^t\gamma_5=\frac{1}{\sqrt{2}}(\gamma^++\gamma^-)\gamma_5,\nonumber\\
\gamma^z\gamma_5=\frac{1}{\sqrt{2}}(\gamma^+-\gamma^-)\gamma_5,
\end{align}
Due to the momentum along the light-cone, operators with $\gamma^-\gamma_5$ correspond to higher order terms of TMDWFs. Therefore, power corrections arising from finite $P^z$ are likely to be eliminated in the average of these two terms:
\begin{align}
\tilde{\Phi}^{\pm}=\frac{1}{2}\left[\tilde{\Phi}^{\pm}\left(\Gamma=\gamma^t\gamma_5\right) + \tilde{\Phi}^{\pm}\left(\Gamma=\gamma^z\gamma_5\right)\right].
\end{align}
For a quantitative analysis see the appendix of~\cite{LatticeParton:2020uhz}. The operator mixing effect reaches order 5\%~\cite{LatticeParton:2020uhz}, which is much smaller than the systematic uncertainties discussed in next subsection. 

According to our numerical simulations, the subtracted quasi TMDWFs in coordinate space $\tilde{\Phi}^{\pm}(z,b_{\perp},P^z)$ as a function of $\lambda=zP^z$ are complex, which is shown in Fig.~\ref{fig:qwf_co}. The examples are the real and imaginary part of $\tilde{\Psi}^{\pm}(x,b_{\perp},P^z)$ with $P^z=24\pi/n_s$, $b_{\perp}=2a$ and $4a$. 
To determine quasi TMDWFs in momentum space $\tilde{\Psi}^{\pm}(x,b_{\perp},P^z)$, we use a  Fourier transformation (FT)
\begin{align}
\tilde{\Psi}^{\pm}(x,b_{\perp},P^z)=\frac{1}{2\pi}\sum_{z_{\rm min}}^{z_{\rm max}}e^{ixzP^z}\tilde{\Phi}^{\pm}(z,b_{\perp},P^z).
\label{eq:FT}
\end{align}
Due to the imaginary part of $\tilde{\Phi}^{\pm}(z,b_{\perp},P^z)$,  $\tilde{\Psi}^{\pm}(x,b_{\perp},P^z)$ also has an imaginary part. We obtain the quasi TMDWFs in momentum space for both real and imaginary part of $\tilde{\Psi}^{\pm}(x,b_{\perp},P^z)$ shown in Fig.\ref{fig:qwf_momenta}, by taking $P^z=24\pi/n_s$ and $b_{\perp}=2a$ and $4a$ as examples. We truncate the FT at $z_{\rm min}$ and $z_{\rm max}$. The deviation of
  $\tilde{\Phi}^{\pm}(z_{\rm min},b_{\perp},P^z)$ and $\tilde{\Phi}^{\pm}(z_{\rm max},b_{\perp},P^z)$ from zero is a measure of the resulting truncation error. For the largest range of $z$ values we could realize numerically, $z_{\rm min}=-1.44$fm, $z_{\rm max}=1.44$fm, this error is still noticeable. This brute-force truncation of the FT leads to an oscillatory behavior of TMDWFs. This oscillation in $\tilde{\Phi}(x,b_{\perp},P^z)$ can be eliminated by an appropriate extrapolation for $\tilde{\Phi}(z,b_{\perp},P^z)$ as a function of $zP^z$ before Fourier transformation. While the signal-to-noise ratio of our data is not smooth enough, the brute-force Fourier transformation is adopted. 
\newpage
\begin{widetext}

\begin{figure}
\centering
\includegraphics[scale=0.55]{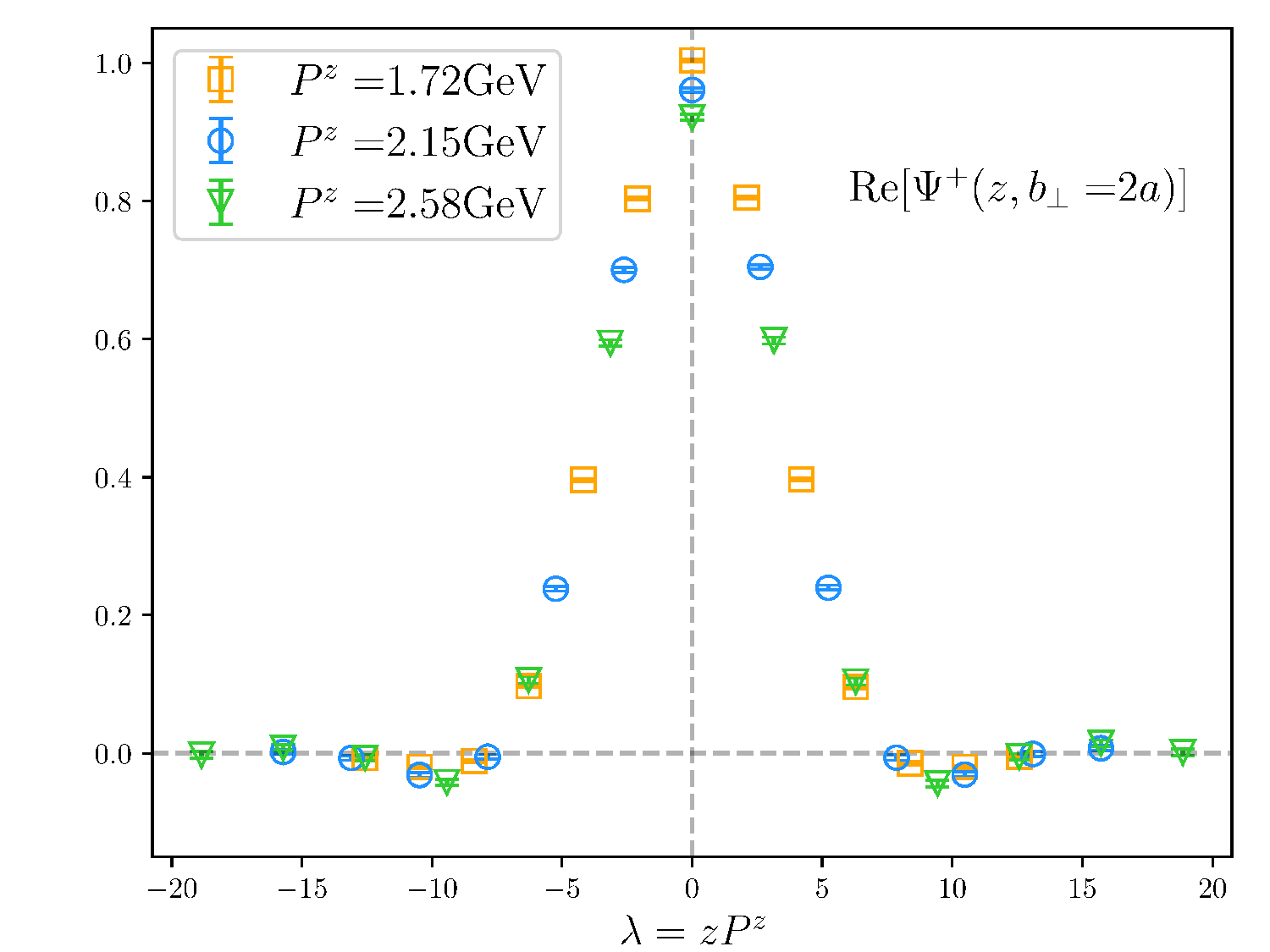}
\includegraphics[scale=0.55]{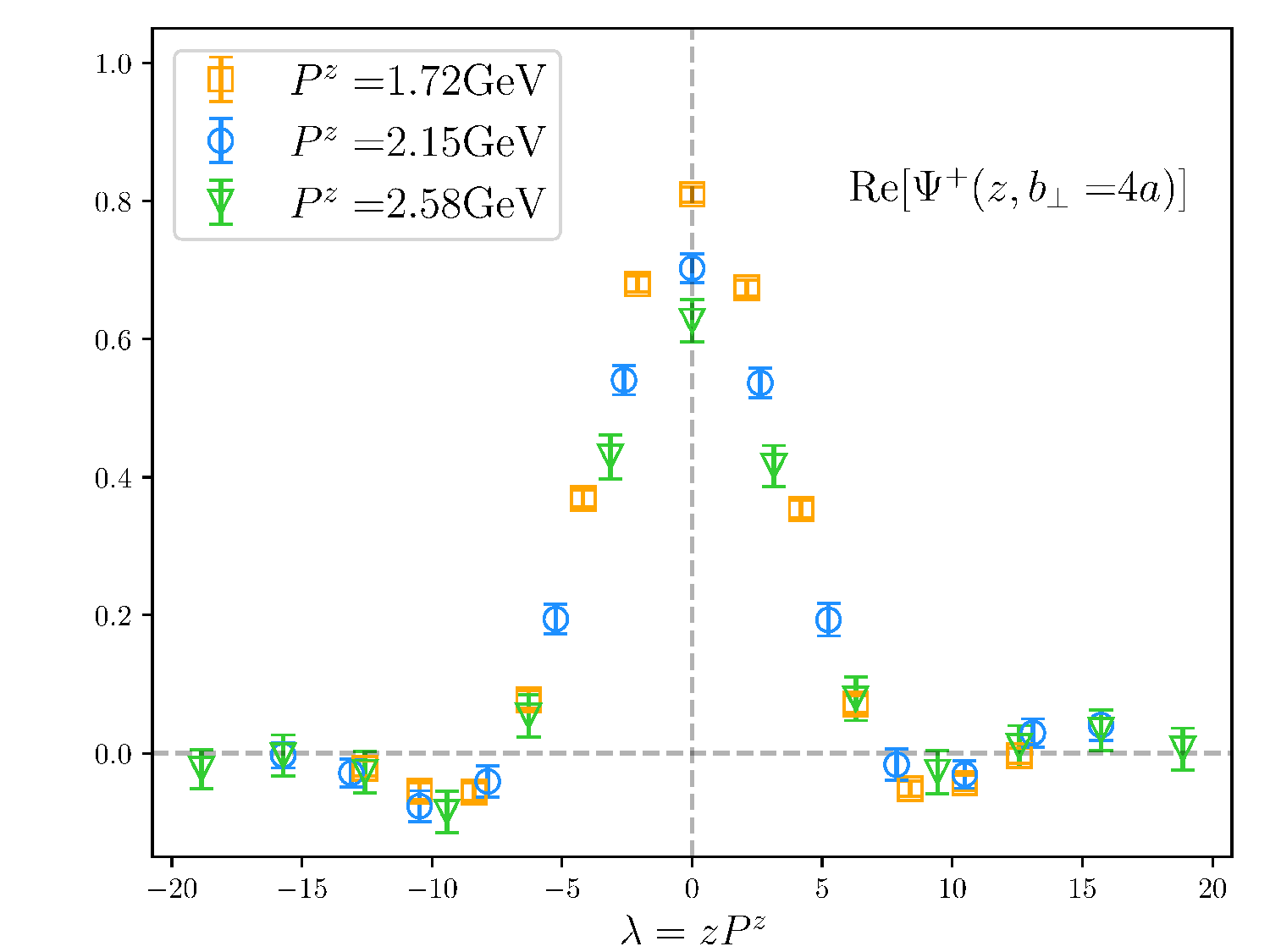}
\includegraphics[scale=0.55]{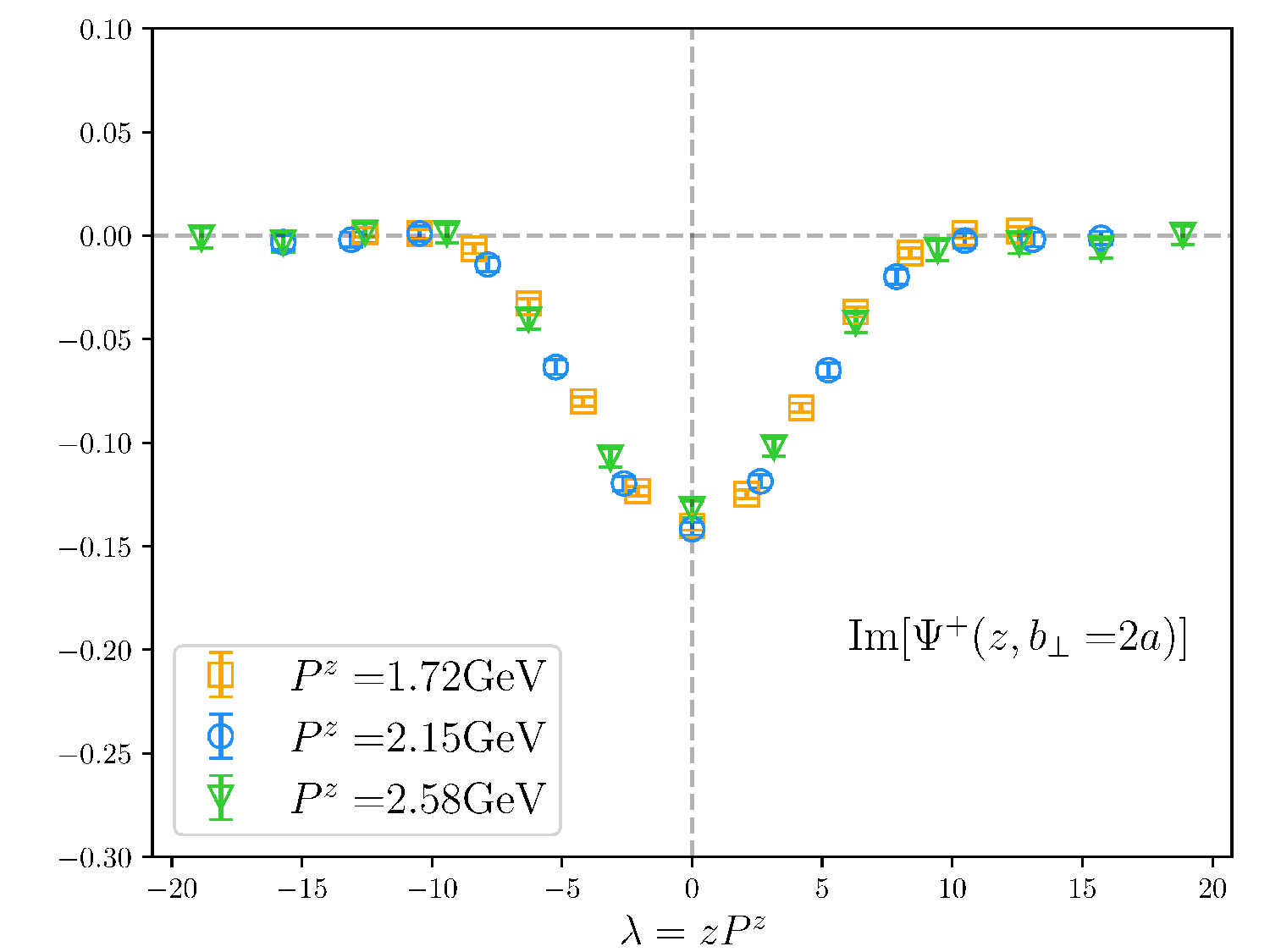}
\includegraphics[scale=0.55]{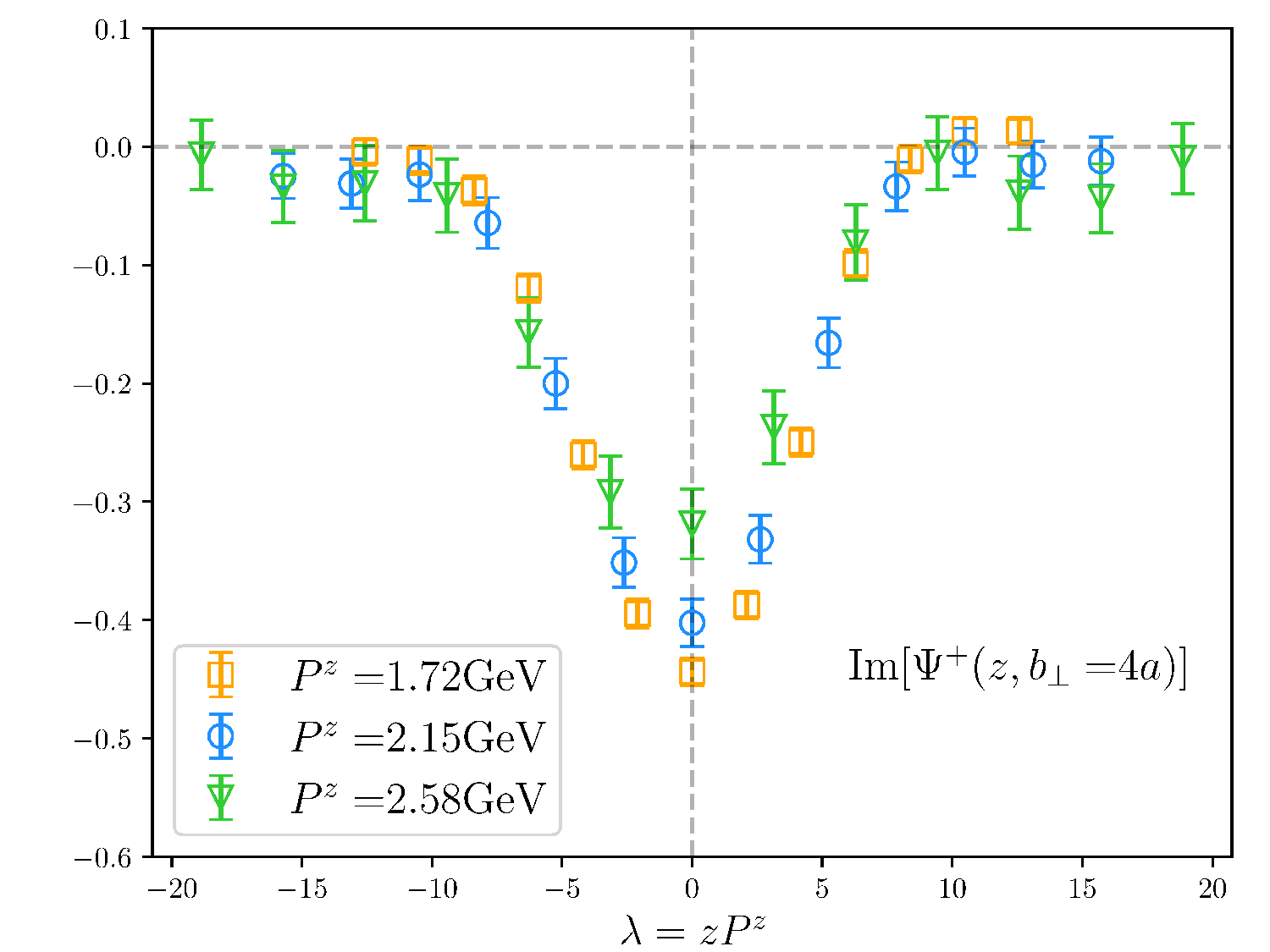}
\caption{Examples for subtracted quasi TMDWFs in coordinate space. Here we take the cases of $P^z=24\pi/n_s$. According to our results, the subtracted quasi TMDWFs $\tilde{\Psi}^+(z,b_{\perp},P^z)$ are complex, thus the real and the imaginary part both need to be investigated. The upper four figures are for subtracted quasi TMDWFs $\tilde{\Psi}^+(z,b_{\perp},P^z)$ as a function of $\lambda=zP^z$ after the average over the two Dirac structures ($\gamma^t\gamma_5$ and $\gamma^z\gamma_5$) is taken. The upper left figure shows the real part of $\tilde{\Psi}^+(z,b_{\perp},P^z)$ with $b_{\perp}=2a$, while the upper right one is for $\tilde{\Psi}^+(z,b_{\perp},P^z)$ with $b_{\perp}=4a$. The lower two figures show the corresponding imaginary parts with $b_{\perp}=2a$ and $4a$.}
    \label{fig:qwf_co}
\end{figure}

\begin{figure}
\centering
\includegraphics[scale=0.55]{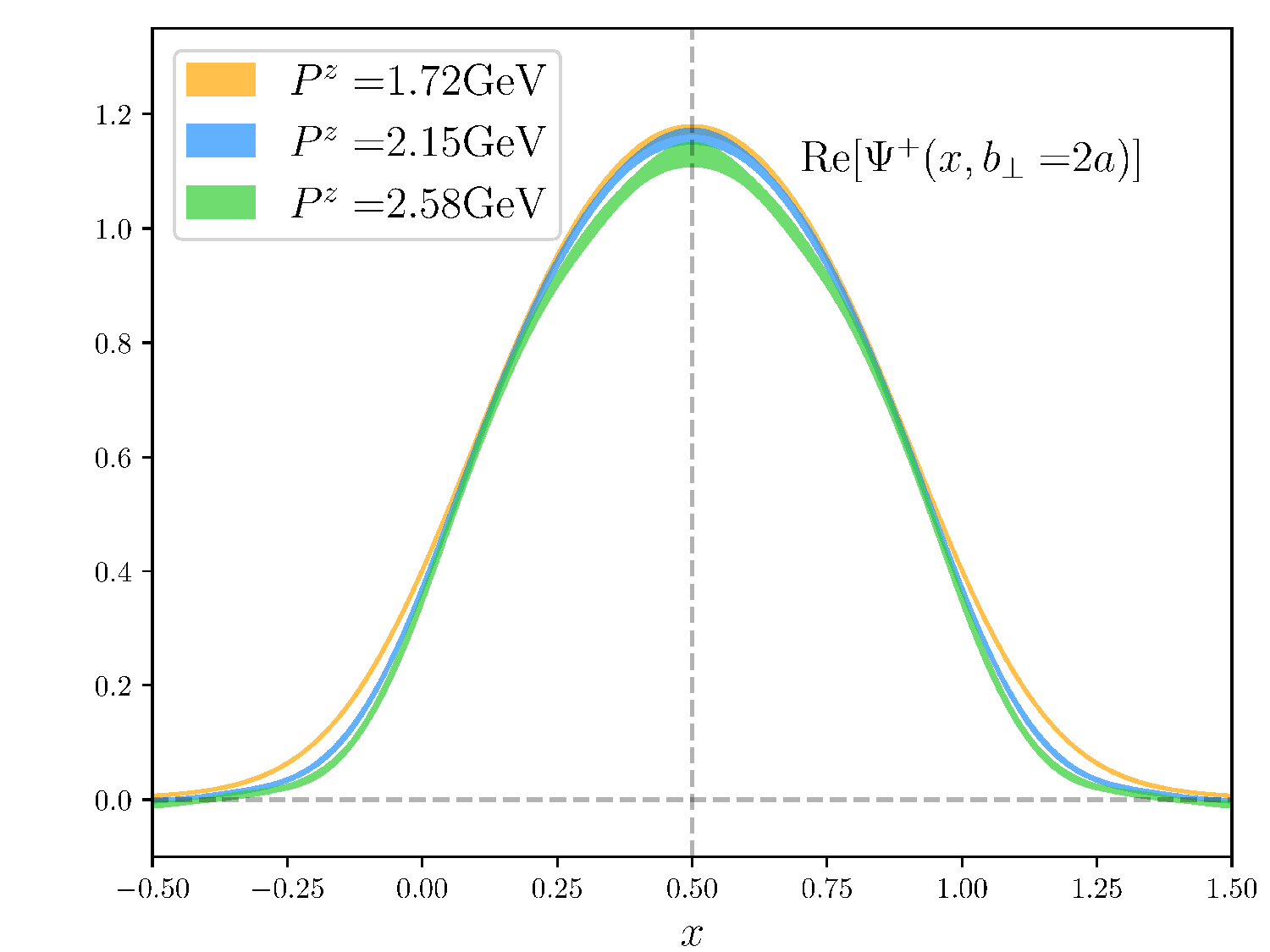}
\includegraphics[scale=0.55]{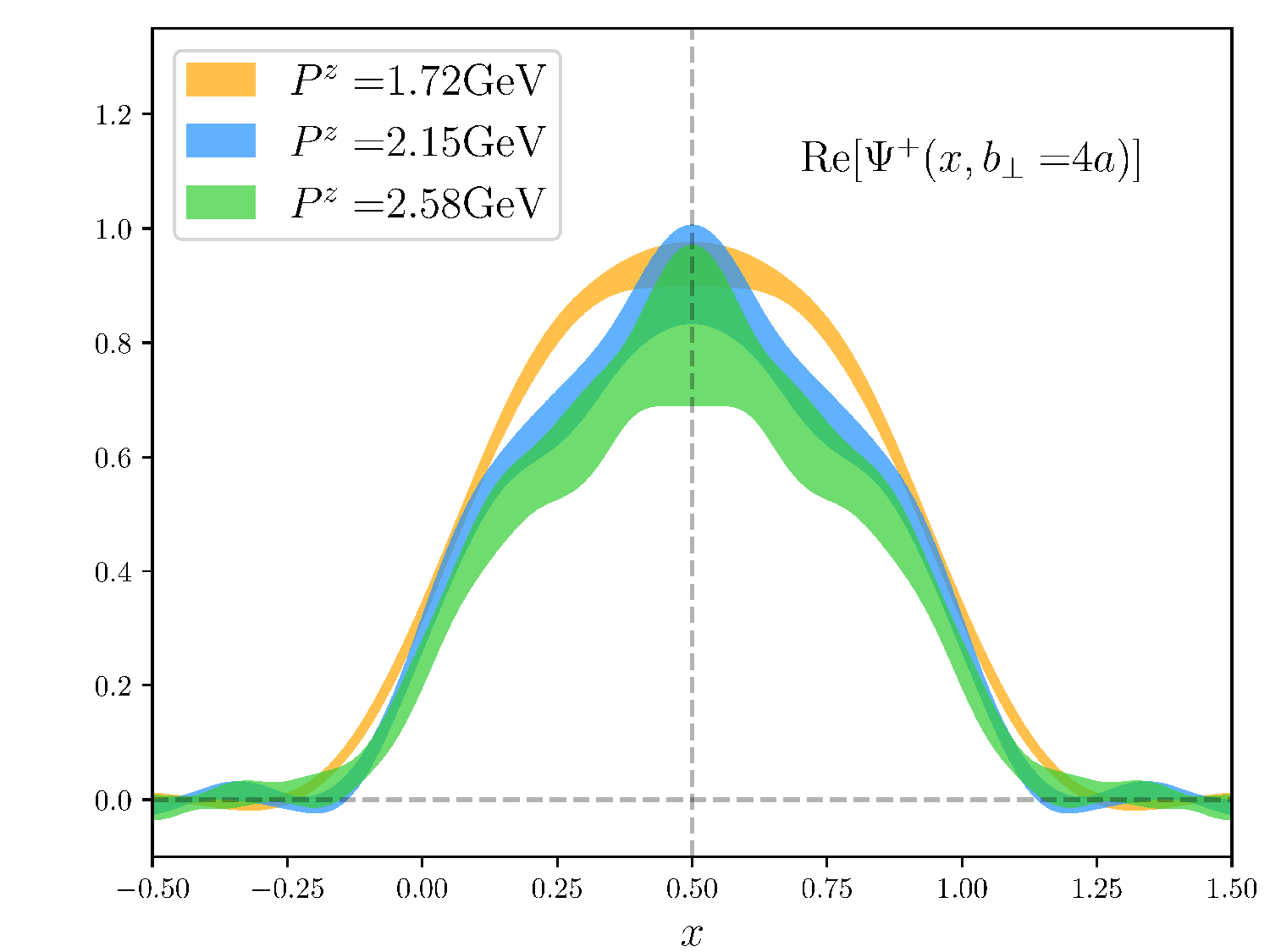}
\includegraphics[scale=0.55]{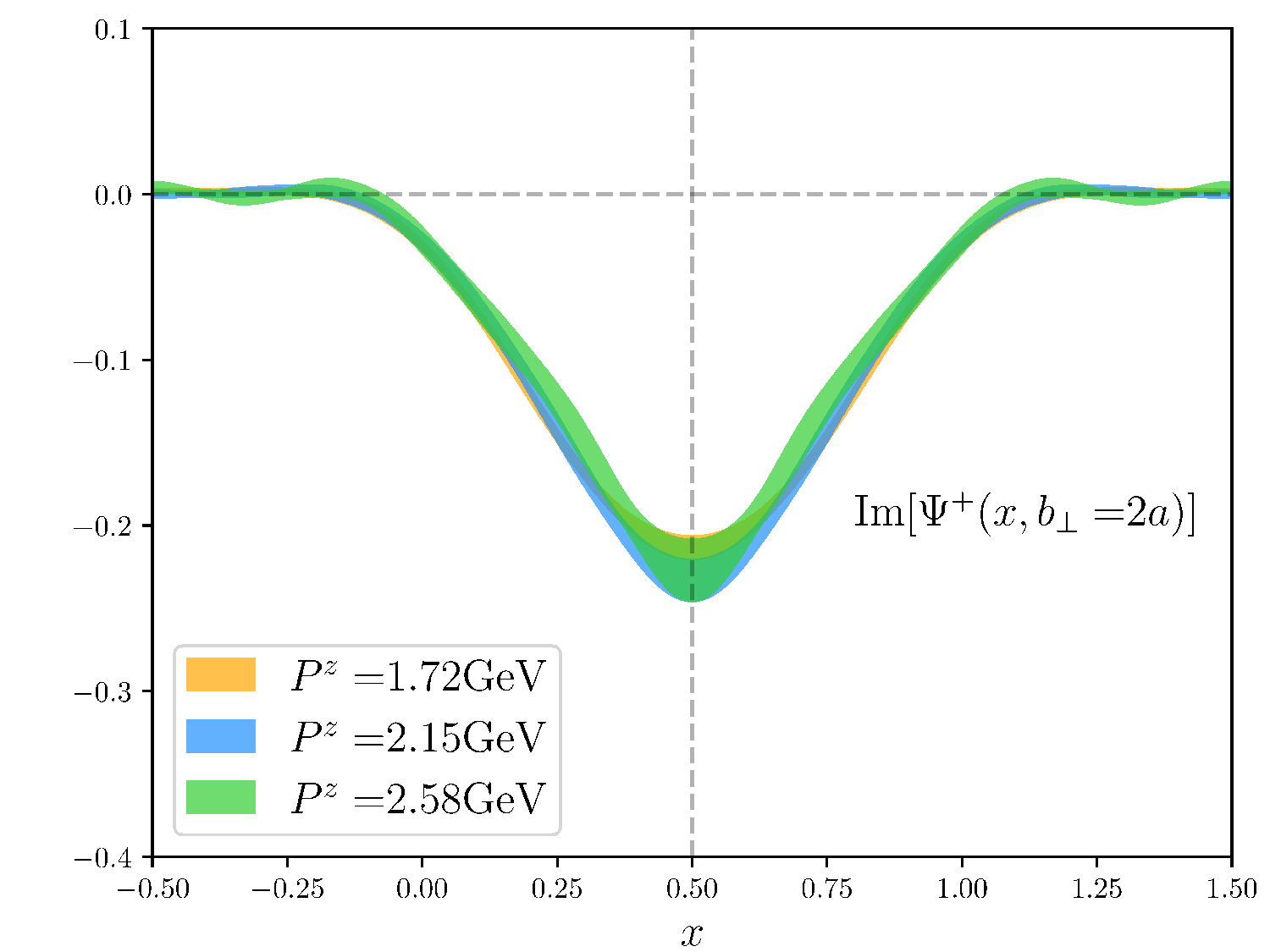}
\includegraphics[scale=0.55]{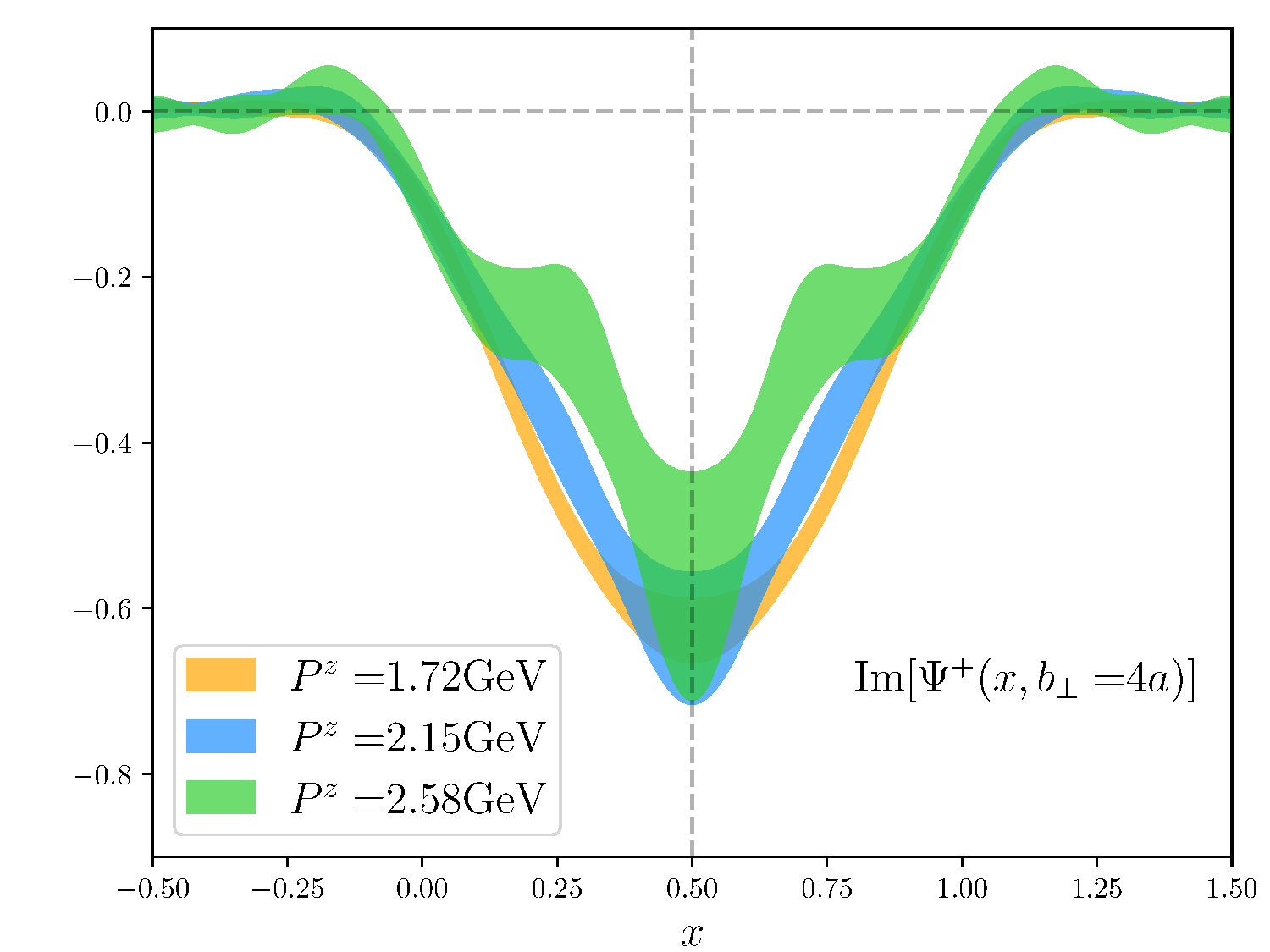}
\caption{Examples for subtracted quasi TMDWFs in momentum space with hadron momentum $P^z=24\pi/n_s$. The subtracted quasi TMDWFs in momentum space $\tilde{\Psi}^+(x,b_{\perp},P^z)$ are Fourier transformed from $\tilde{\Phi}^+(z,b_{\perp},P^z)$, which have real and imaginary part. Both have to be investigated. Shown in Eq. (\ref{eq:FT}), a brute-force FT is used to determine $\tilde{\Psi}^+(x,b_{\perp},P^z)$, where $z_{\rm min}=-1.44$~fm, $z_{\rm max}=1.44$~fm. The upper left figure shows the real part for $\tilde{\Psi}^+(x,b_{\perp},P^z)$ with $b_{\perp}=2a$ case, while the upper right one is for $\tilde{\Psi}^+(x,b_{\perp},P^z)$ with $b_{\perp}=4a$ case. Correspondingly, the lower two figures shows the imaginary part for $\tilde{\Psi}^+(x,b_{\perp},P^z)$ with $b_{\perp}=2a$ and $4a$.}
    \label{fig:qwf_momenta}
\end{figure}

\end{widetext}

\subsection{Collins-Soper Kernel From Quasi TMDWFs}
\label{subsec:csresults}

The CS kernel governs the rapidity evolution and thus is independent of the momentum fraction of the involved parton.  But as indicated in Eq.(\ref{eq:matching}), the factorization formula  works only when $xP^z\gg \Lambda_{\rm QCD}$, and could be invalid  in the end-point regions $x\to0,1$. Power corrections are presumably  of the form $1/\left(xP^z\right)^2$ or $1/\left(\bar{x}P^z\right)^2$. Therefore,  the numerical CS kernel is fitted by a function of $x$, $P_1^z$ and $P_2^z$ and is written as $K(b_{\perp},\mu,x,P^z_1,P^z_2)$,
\begin{align}
&K(b_{\perp},\mu,x,P^z_1,P^z_2) =\nonumber\\
&\frac{1}{2\ln(P_1^z/P_2^z)} \left[\ln\frac{H^{+}(xP_2^z,\mu)\tilde{\Psi}^{+}(x,b_{\perp},\mu,P_1^z)}{H^{+}(xP_1^z,\mu)\tilde{\Psi}^{+}(x,b_{\perp},\mu,P_2^z)}\right.  \nonumber\\
	&+\left.\ln\frac{H^{-}(xP_2^z,\mu)\tilde{\Psi}^{-}(x,b_{\perp},\mu,P_1^z)}{H^{-}(xP_1^z,\mu)\tilde{\Psi}^{-}(x,b_{\perp},\mu,P_2^z)}\right].
\end{align}
Here $K(b_{\perp},\mu,x,P^z_1,P^z_2)$ are extracted from the perturbative matching kernels and quasi TMDWFs using 1-loop matching. They will have power corrections of teh form $\mathcal{O}\left(1/\left(xP^z\right)^2\right)$ and $\mathcal{O}\left(1/\left(\bar{x}P^z\right)^2\right)$. In order to extract the leading power contributions, we adopt  the following parametrization
\begin{align}
&K(b_{\perp},\mu,x,P^z_1,P^z_2) = K(b_{\perp},\mu) \nonumber\\
&\quad+A\left[\frac{1}{x^2(1-x)^2(P^z_1)^2}-\frac{1}{x^2(1-x)^2(P^z_2)^2}\right], \label{eq:parametrizationofCSkernel}
\end{align}
where $A$ is the coefficient accounting for  the leading higher power contributions, and can be determined through a joint fit of different lattice data in the regions not so close to $x=0, 1$. 

\begin{widetext}

\begin{figure}
\centering
    \includegraphics[scale=0.55]{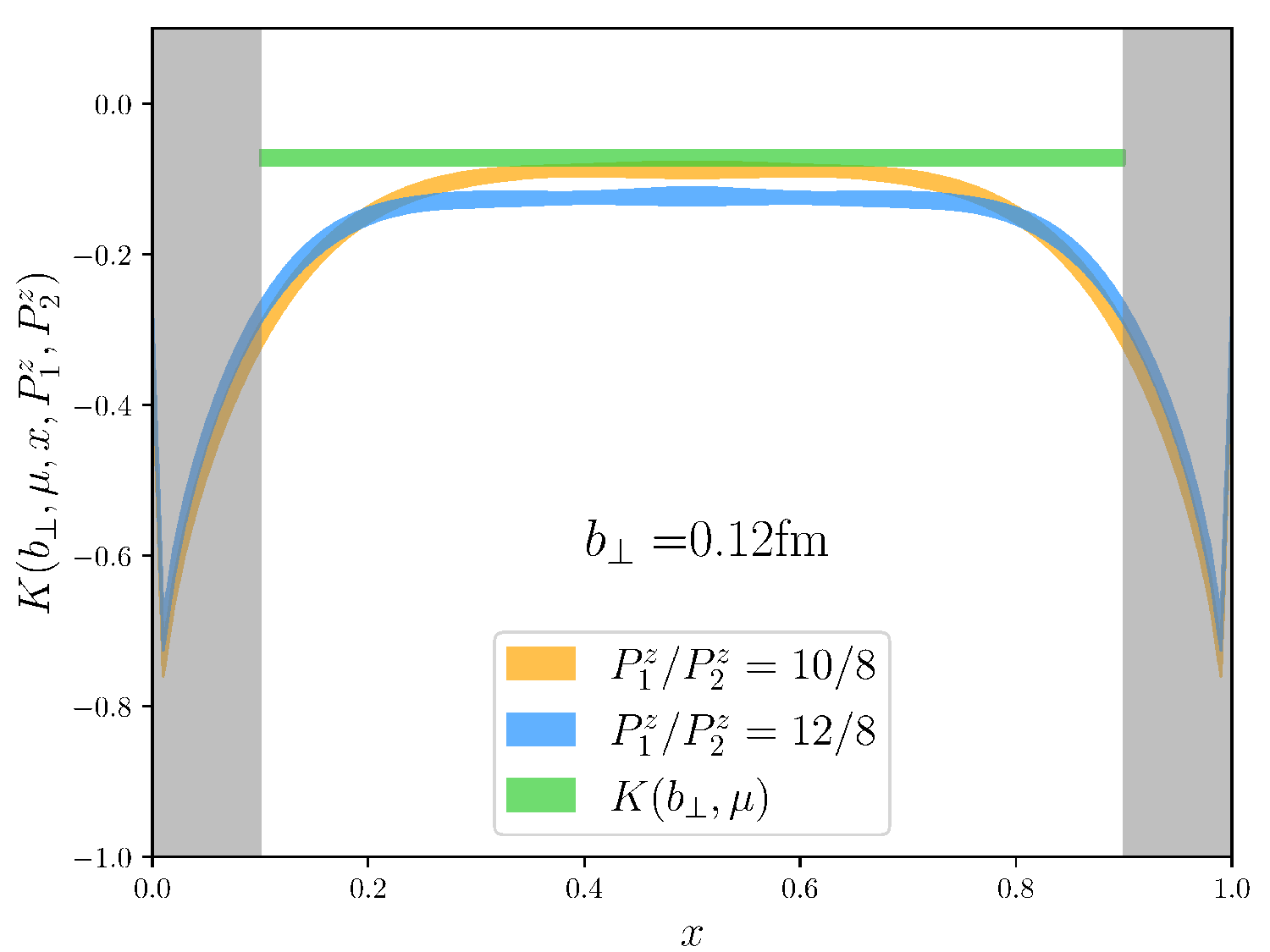}
    \includegraphics[scale=0.55]{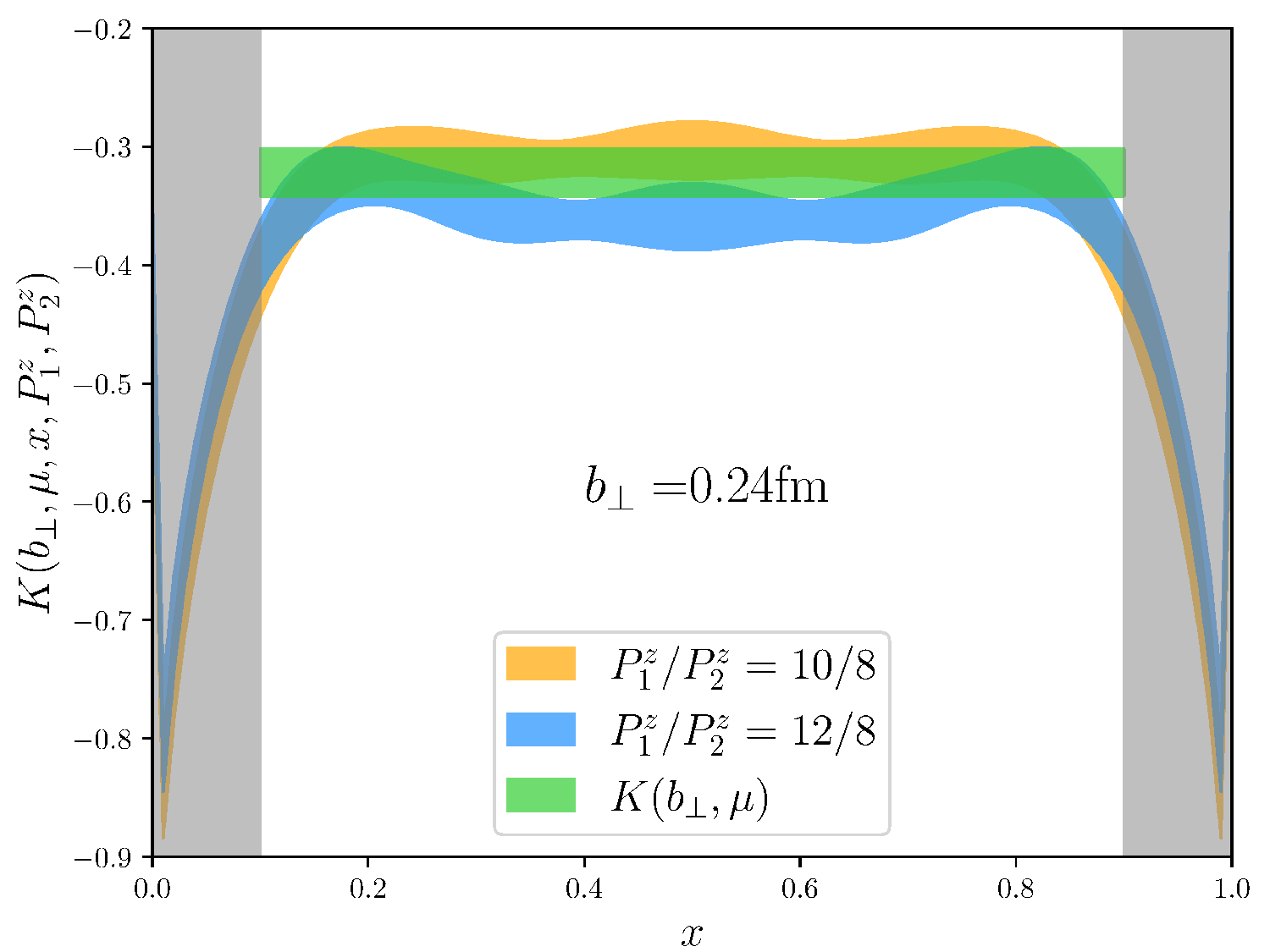}
    \includegraphics[scale=0.55]{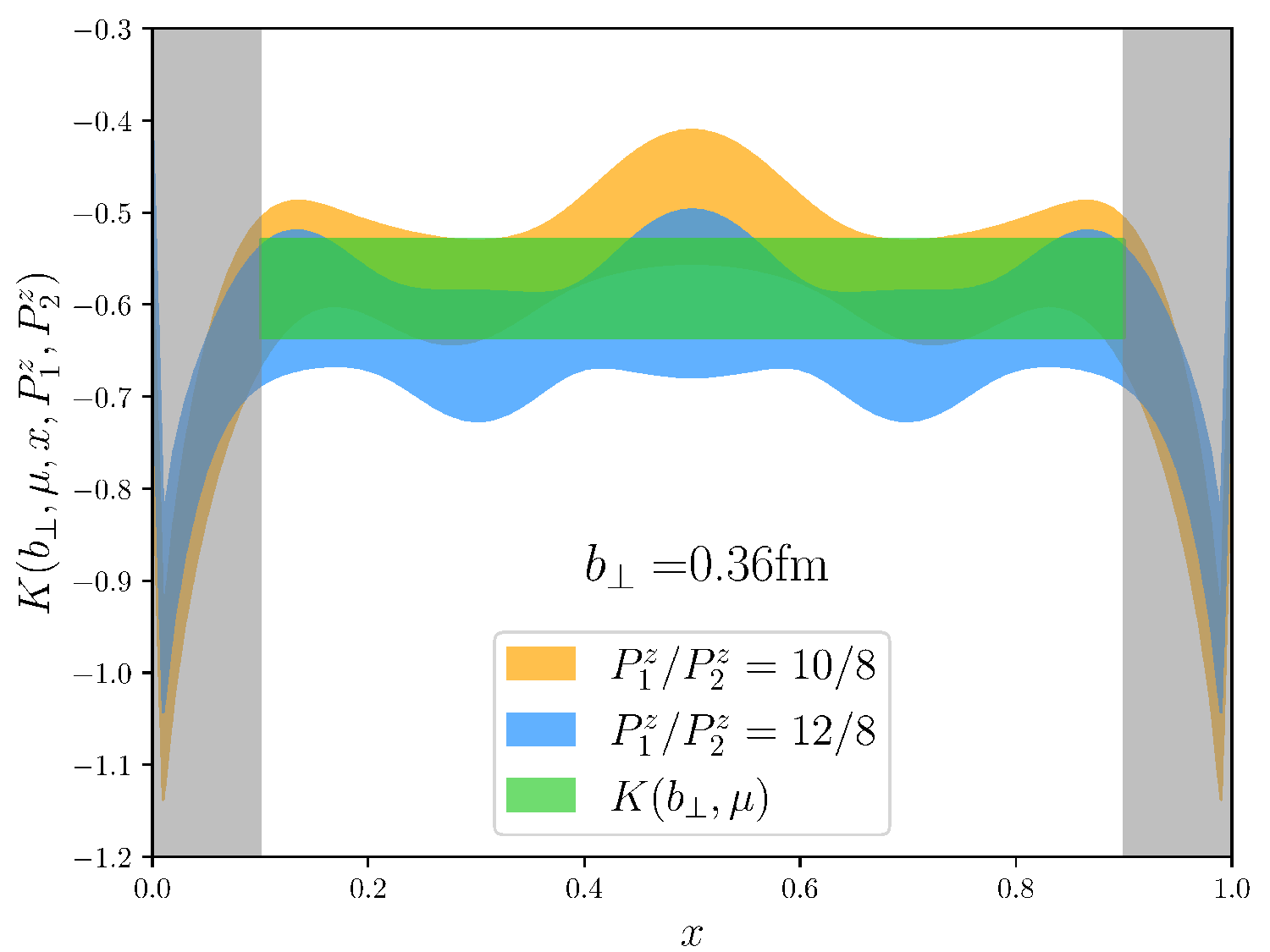}
    \includegraphics[scale=0.55]{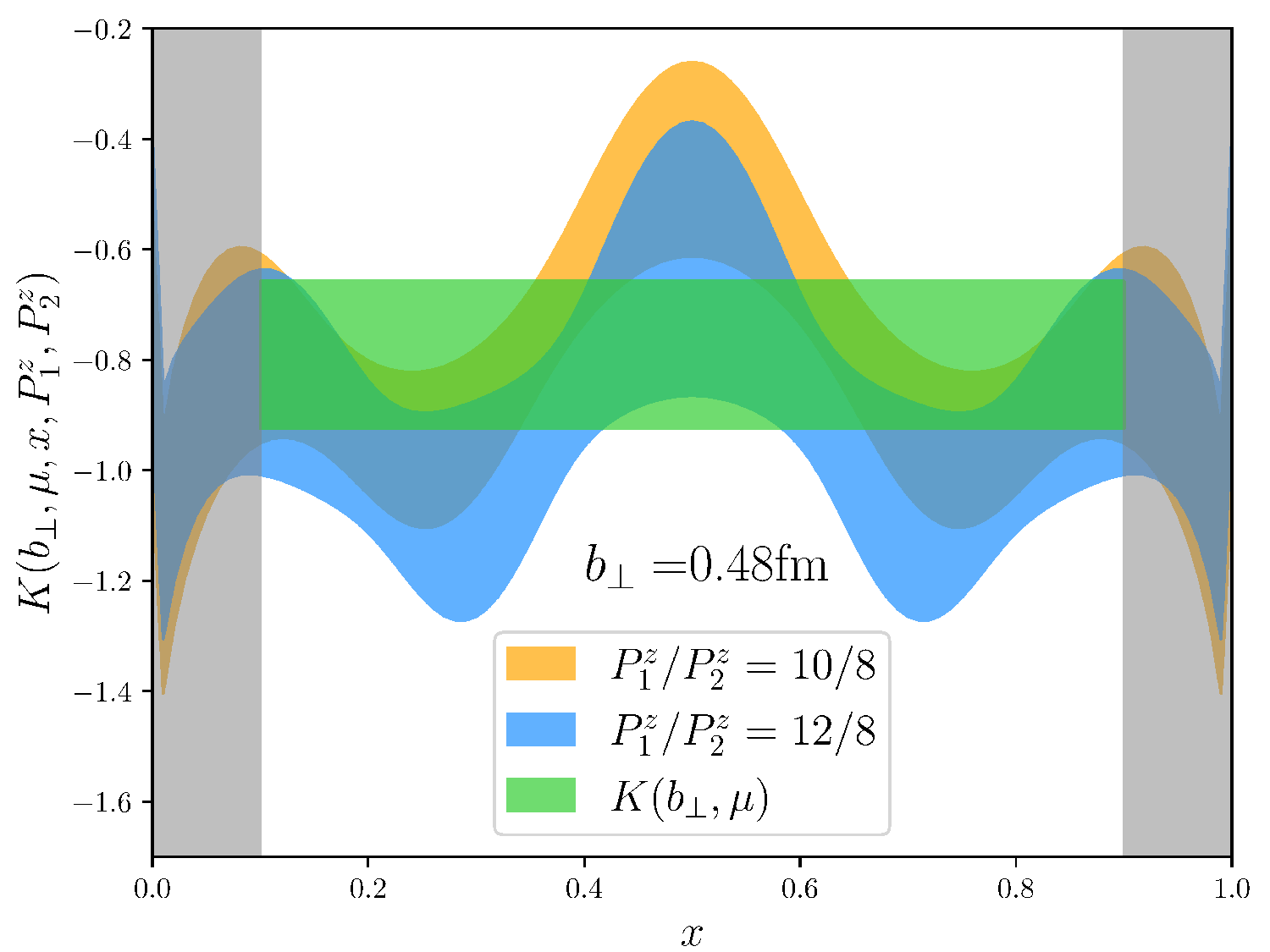}
    \includegraphics[scale=0.55]{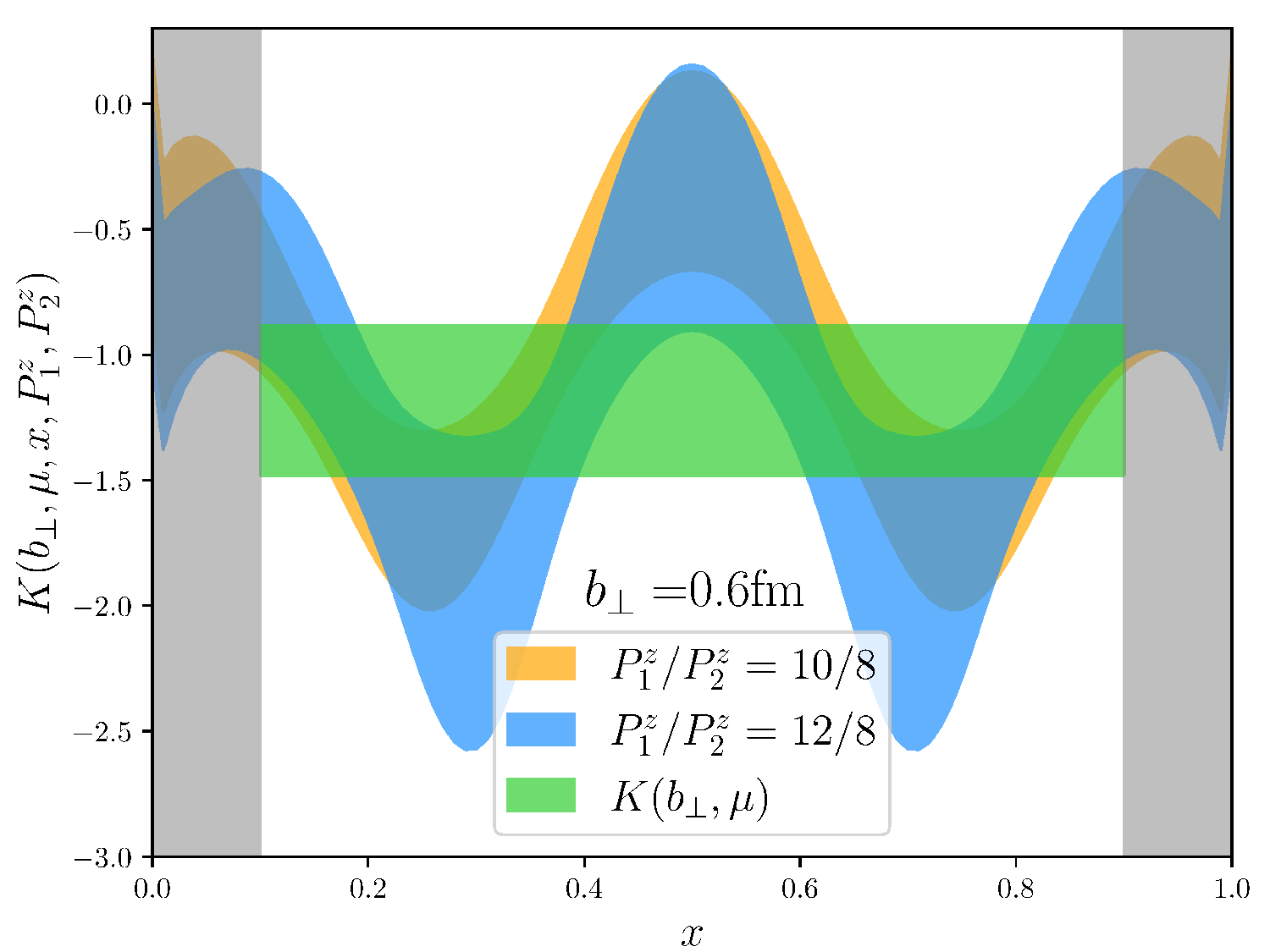}
\caption{The fit results of $K(b_{\perp},\mu,x,P^z_1,P^z_2)$ extracted from the quasi-WFs $\tilde{\Psi}^{\pm}$. The chosen momentum pairs $\{P_1^z,P^z_2\}$ are denoted by $P^z_1/P_2^z$ in the legend. The figures are corresponding to cases with $b_{\perp}=\{0.12,0.24,0.36,0.48,0.60\}$fm.} The horizontal shaded band shows the central value and uncertainty of $K(b_{\perp},\mu)$, as well as the fit range of $x$, as described in the text. In large $b_{\perp}$ area, the strong oscillation exists in the shaded area at both edges, where LaMET approach is invalid, is caused by the breakdown of the large momentum expansion.
    \label{fig:K_b}
\end{figure}

\end{widetext}

Fig.~\ref{fig:K_b} presents the physical CS kernel $K(b_{\perp},\mu)$ with $b_{\perp}=\{0.12,0.24,0.36,0.48,0.60\}$fm. By employing three cases of quasi TMDWFs with $P^z=\{8,10,12\}\times2\pi/n_s$, one can extract $K(b_{\perp},\mu,x,P^z_1,P^z_2)$ with $P_1^z/P_2^z=10/8$ and $12/8$, shown as the different colored bands.  Except in the end-point regions ($x<0.2$ or $x>0.8$), the lattice data is flat and reflects the leading power contribution, which conforms with expectations. Using the parametrization formula Eq.(\ref{eq:parametrizationofCSkernel}), the physical CS kernel $K(b_{\perp},\mu)$ can be determined by fitting the data, shown as the green band. As mentioned before, at large $b_{\perp}$, the quasi TMDWFs   show oscillations due to the  truncation of the Fourier transformation, which also affect the extracted $K(b_{\perp},\mu,x,P^z_1,P^z_2)$, as shown in the lower panel of Fig.\ref{fig:K_b}. This oscillation effect can be in principle removed once larger $z$ data becomes possible, or if one knows how to extrapolated the current data to the larger $z$ or $\lambda$ region. 

\begin{figure}
\centering
\includegraphics[scale=0.55]{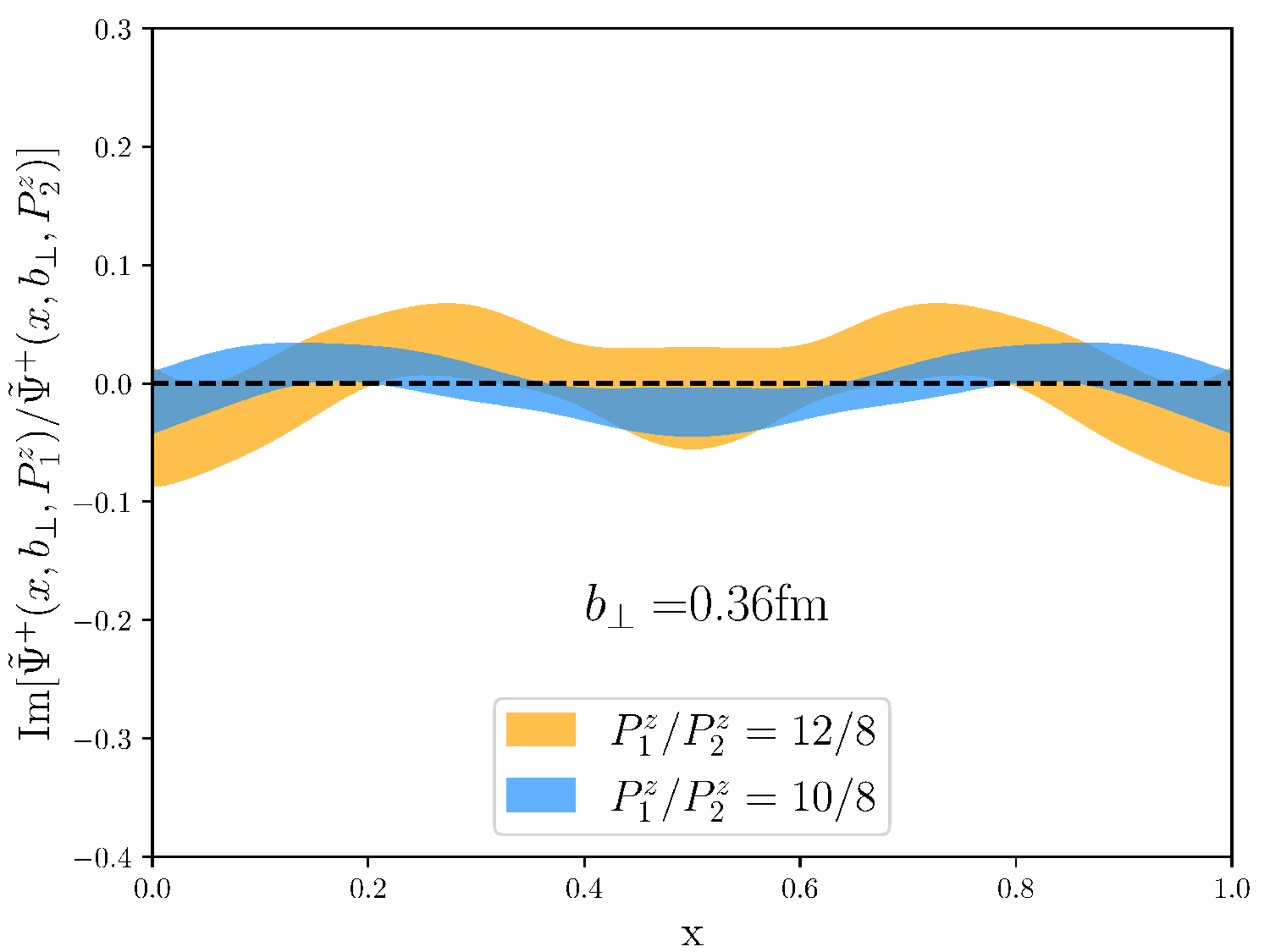}
\caption{Example of numerical results for the imaginary part of the ratio for quasi TMDWFs at different $P^z$, $\frac{\tilde{\Psi}^+(x,b_{\perp},\mu,P_1^z)}{\tilde{\Psi}^+(x,b_{\perp},\mu,P_2^z)}$, with $b_{\perp}=3$a as a function of momentum fraction $x$. The imaginary parts of both cases are close to zero.}
    \label{fig:ratio_phi}
\end{figure}

Theoretically the physical CS kernel is purely real, however, there still exists a residual imaginary part at 1-loop matching. As discussed above, this imaginary term comes from the perturbative matching kernel. It is easily to prove that $H^{\pm}(zP_2^z,\mu)/H^{\pm}(zP_1^z,\mu)$ contains imaginary part while the lattice results for $\tilde{\Psi}^{\pm}(x,b_{\perp},\mu,P_1^z)/\tilde{\Psi}^{\pm}(x,b_{\perp},\mu,P_2^z)$ are nearly real, shown as \ref{fig:ratio_phi}. Therefore, we consider this imaginary part as systematic uncertainty from factorization theorem. This uncertainty can be expressed as

\begin{align}
&\sigma_{\mathrm{sys}}=\sqrt{K(b_{\perp},\mu)+\text{Im}^2\left[K^+(b_{\perp},\mu)\right]}-K(b_{\perp},\mu)
\label{eq:systematical_error}
\end{align}
where $\text{Im}\left[K^+(b_{\perp},\mu)\right]$ represents the numerical imaginary part of extracting $K(b_{\perp},\mu)$ only by $\tilde{\Psi}^+$:
\begin{align}
K^+(b_{\perp},\mu)=&\frac{1}{\ln (P^z_1/P^z_2)}\ln\frac{H^+(xP^z_2,\mu)\tilde \Psi^+(x, b_\perp, \mu, P^z_1)}{H^+(xP^z_1,\mu)\tilde \Psi^+(x, b_\perp, \mu, P^z_2)}.
\label{eq:K+}
\end{align}

It should be noticed that the perturbative matching kernel $H^+(zP^z,\mu)$ is the complex conjugate of $H^-(zP^z,\mu)$, that is the imaginary parts in these terms can be cancelled each other when we employ the average of $H^+(zP_2^z,\mu)/H^+(zP_1^z,\mu)$ and $H^-(zP_2^z,\mu)/H^-(zP_1^z,\mu)$. Therefore, as the final result, we adopt $K(b_{\perp},\mu)=\left[K^+(b_{\perp},\mu)+K^-(b_{\perp},\mu)\right]/2$ to reserve the real part, and regard the imaginary contributions as our systematic uncertainty.

\subsection{Results and Discussions}

One should notice that the Wilson loop renormalized quasi TMDWF on the lattice (Eq.(\ref{eq:WLRQuasi})) has a scale dependence on $a$. If one converts it to the $\overline{\rm MS}$ scheme through dividing it by $Z_{O}$ (Eq.(\ref{eq:ZO_RS_higher})), the scale $\mu$ is introduced. In principle, one should convert the Wilson loop renormalized quasi TMDWF to the $\overline{\rm MS}$ scheme since our factorization formula works there. However, since $Z_{O}$ has no dependence on momentum $P_{z}$, it cancels in the ratio of quasi TMDWFs, so does the scale dependence. So, one does not need to do the scheme and scale conversion of the quasi TMDWF during the extraction of CS kernel.

\begin{figure}
\centering
\includegraphics[scale=0.55]{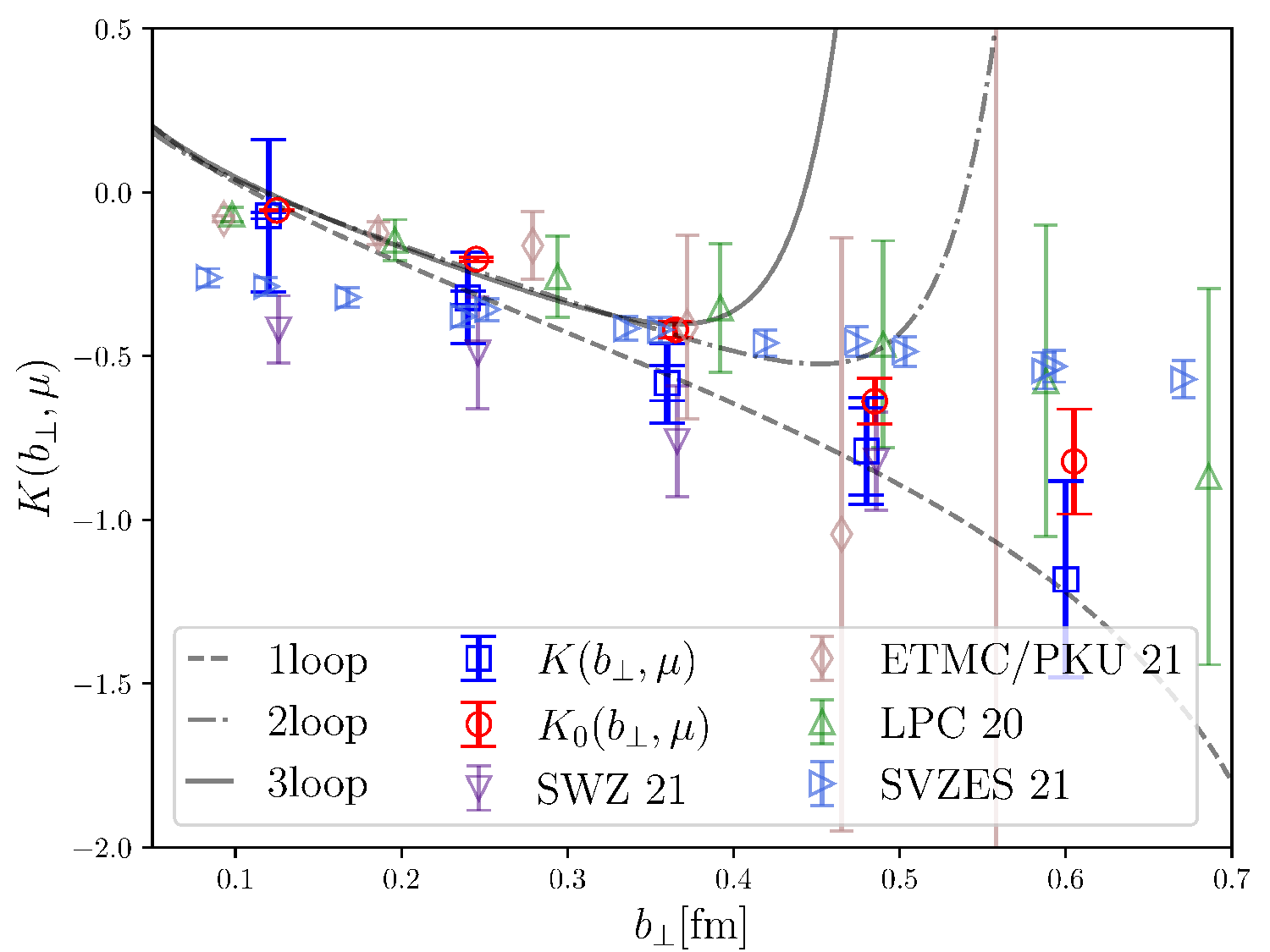}
\includegraphics[scale=0.55]{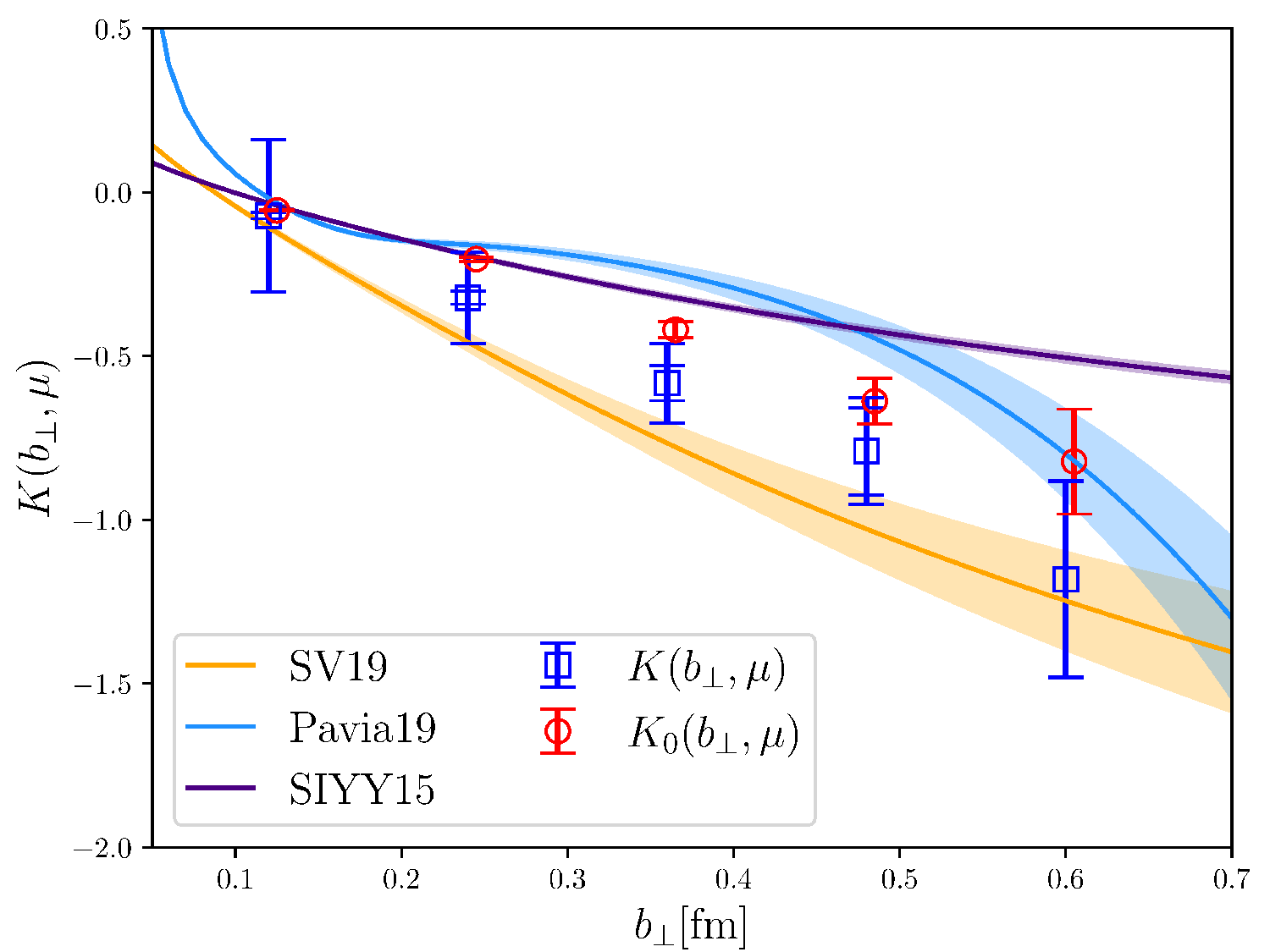}
\caption{{The upper panel shows the comparison of our results $K(b_{\perp},\mu)$ and $K_0(b_{\perp},\mu)$ with the lattice calculations by SWZ~\cite{Shanahan:2021tst}, LPC~\cite{LatticeParton:2020uhz}, ETMC/PKU~\cite{Li:2021wvl} and SVZES\cite{Schlemmer:2021aij}, as well as the perturbative calculations up to 3-loop. $K(b_{\perp},\mu)$ denotes the CS kernel extracted through 1-loop matching, and whose uncertainties correspond to the statistical errors and the systematic ones from the non-zero imaginary part. $K_0(b_{\perp},\mu)$ denotes our tree-level results, only with statistical uncertainties. The lower panel shows the comparison of our result with phenomenological extractions: SV19~\cite{Scimemi:2019cmh}, Pavia19~\cite{Bacchetta:2019sam} and SIYY15~\cite{Sun:2014dqm} give phenomenological parameterizations of CS kernel fitted to data from} high energy collision processes like Drell-Yan.}
\label{fig:K_com}
\end{figure}

The extracted CS kernel from the combined fit of the ratios of quasi TMDWFs  with different momenta are shown as the red data points in Fig. \ref{fig:K_com}. {In this figure we exhibit two kinds of errors for $K(b_{\perp},\mu)$, in which the smaller ones denote statistical uncertainties while the larger ones include both statistical and systematical uncertainties}. In the small-$b_{\perp}$ region, systematical uncertainties are dominant due to the large power and the nonzero imaginary part. 

As a comparison, we also give the tree-level matching  result for the CS kernel.  With the leading order matching kernel $H(xP^z,\mu)=1+\mathcal{O}(\alpha_s)$, Eq. (\ref{eq:extractingCSkernel}) simplifies to the ratio of quasi TMDWFs $\tilde{\Phi}$ at $z=0$ with momentum $P_1^z/P^z_2$. The blue dots in Fig.~\ref{fig:K_com} denote the results obtained for tree-level matching, for which only statistical uncertainties are shown. 

We compare our results with the ones from perturbative calculations, phenomenological extractions as well as the lattice results determined by other collaborations. 

The black solid and dashed lines in the upper panel of Fig.~\ref{fig:K_com} indicate the perturbative results  up to 3-loops, with a running coupling constant $\alpha_s(\mu=1/b_{\perp})$. The perturbative calculations work well in small $b_{\perp}$ region ($b_{\perp}\ll1/\Lambda_{\mathrm{QCD}}$), while will diverge with $b_{\perp}$ increasing. In contrast, the lattice calculation will give accurate predictions in the nonperturvative region, while due to the power corrections, it 
might suffer large systematic uncertainties in small $b_{\perp}$ region.

Similar to this work, the results of LPC~\cite{LatticeParton:2020uhz} and ETMC/PKU~\cite{Li:2021wvl} are also extracted from quasi TMDWFs through a tree-level matching. Adopting the one-loop matching formula, as well as considering the operator mixing effects will help ones to reduce the systematic uncertainties and obtain more precise results. In addition, considering the different directions of gauge link will help us to eliminate the contributions from unphysical imaginary part, and then improve the accuracy of our results.

In another way, the SWZ~\cite{Shanahan:2021tst} and SVZES\cite{Schlemmer:2021aij} results are obtained from quasi-TMDPDFs. Compared with the complicated nucleon correlation functions, the meson ones are much easier to obtain better signals.  Besides, the wave functions of meson are nearly symmetric in $x$-space, it thereby more convenient to parametrize the oscillation effects and obtain the physical results in large $P^z$ limit like Fig.(\ref{fig:K_b}). In addition, the light meson is more easily to reach a larger boosted factor, one can see from the small $b_{\perp}$ region, the results from quasi TMDWFs fit well with the perturbative calculations than the ones from quasi TMDPDFs.

The lower panel of Fig.(\ref{fig:K_com}) shows the comparison with phenomenological results. SV19~\cite{Scimemi:2019cmh} and SIYY15~\cite{Sun:2014dqm} use a parameterization with perturbative and nonperturbative parts. However Pavia19~\cite{Bacchetta:2019sam} obtained their result with the factorization of TMDPDFs, obtaining the CS kernel from the rapidity derivative. In addition they fit parameters from the Drell-Yan data to obtain their phenomenological CS kernel. The results from different methods exhibit obviously inconsistency in the nonperturbative region. Our result shows a better consistency with SV19.

\section{Summary and Outlook}
\label{sec:summary}

In this work, we have calculated the CS kernel on a MILC lattice configuration in the large momentum effective theory framework. Comparing with our previous studied~\cite{LatticeParton:2020uhz}, the one-loop matching kernel has been adopted in this study, and several hadron momenta were used to extract the CS kernel. We found that in the small $b_{\perp}$ region, our results are consistent with perturbative QCD. 
In large $b_{\perp}$ region, our results seem consistent with other lattice calculations in the literature within uncertainties. 

For our future studies, we need to
use lattice configurations with multiple lattice spacing to understand the finite
lattice spacing effects. We would
use a valence quark mass consistent with the sea quark one to reduce the non-unitarity effects. One such effect 
might be the imaginary part of the meson
wave function which seems inconsistent
with perturbative calculation at present time. Clearly, all these explorations will
take more computational resources.

\section*{Acknowledgement}
We thank Xu Feng, Yizhuang Liu, and Feng Yuan for useful discussions.  This work is supported in part by Natural Science Foundation of China under grant No. 11735010, 11911530088, U2032102, 11653003, 11975127, 11975051, 12005130, 12147140. MC, JH, WW is also supported  by Natural Science Foundation of Shanghai under grant No. 15DZ2272100.  PS is also supported by Jiangsu Specially Appointed Professor Program. YBY is also supported by the Strategic Priority Research Program of Chinese Academy of Sciences, Grant No. XDB34030303, XDPB15. AS, PS, WW, YBY and JHZ are also supported by the NSFC-DFG joint grant under grant No. 12061131006 and SCHA~~458/22. XJ is partially supported by the U.S. Department of Energy under Contract No. DE-SC0020682. The calculation was supported by Advanced Computing East China Sub-center and the $\pi$2.0 cluster at Center for High Performance Computing, Shanghai Jiao Tong University.

\appendix

\section{Euclidean Time $t$ Dependence of Normalized $C_2$}
\label{appsec:C2}

In Sec. \ref{sec:quasi-WF_from_2pt} the ratio of nonlocal and local  two-point functions is parametrized in Eq.(\ref{eq:two-state-fit})
\begin{align}
&R^{\pm}\left(z,b_{\perp},P^z,L,t \right)=\frac{C_2^{\pm}\left(z,b_{\perp},P^z,L,t \right)}{C_2\left(0,0,P^z,0,t \right)} \nonumber\\
=&\tilde{\Phi}^{\pm0}\left(z,b_{\perp},P^z,L\right)\left[1+c_0\left(z,b_{\perp},P^z,L\right)e^{-\Delta Et} \right].
\end{align}
From the above equation, one can see that $R^{\pm}\left(z,b_{\perp},P^z,L,t \right)$ decays exponentially with $t$. As discussed in Sec. \ref{sec:quasi-WF_from_2pt}, the one-state and two-state fits are both used to extract $\tilde{\Phi}^{\pm0}\left(z,b_{\perp},P^z,L\right)$. As shown in Fig.~\ref{fig:t_dep}, for the cases with small $\{z,b_\perp\}$  as $\{0a,1a\},\{0a,3a\},\{2a,2a\},\{2a,3a\}$, the two-state fit results are consistent with one-state ones. However, for the cases with large $\{z, b_\perp\}$, the excited state contamination can not be well described with two-state parametrization. Since the excited state contamination will decrease with the Euclidean time separation increasing, we use the plateau at large $t$ for our one-state fit.

\section{Gauge-Link Length $L$ Dependence of Quasi TMDWFs}
\label{sec:gauge_line}

In Sec. \ref{subsec:numerical_wl}, the Wilson loop is used to renormalize quasi TMDWFs, which removes the linear divergence. Similar with the discussion in Sec.~\ref{subsec:numerical_wl}, we give results with the different $\{P^z,b_{\perp},z\}$ in Fig.~\ref{fig:multi_l_dep} to show separately the Wilson-link length $L$-dependence of Wilson loop, unsubtracted quasi TMDWFs and subtracted quasi TMDWFs. At large $L$, $\tilde{\Phi}^{+0}$ decays at the same speed of $\sqrt{Z_E}$, so Wilson loop cancels the linear divergence in un-subtracted quasi TMDWFs.

\section{Power Correction Effects}
\label{sec:app_operator_mixing}

As described in Sec. \ref{sec:operator}, quasi TMDWFs for a pseudoscalar meson require  the projectors $\Gamma=\gamma^t\gamma_5$ or $\Gamma=\gamma^z\gamma_5$ onto the leading twist light-cone distribution amplitude, {\it i.e.} $\gamma^+\gamma_5$ in large $P^z$ limit. Fig. \ref{fig:multi_qwf_co} shows examples with different $\{P^z,b_{\perp}\}$ for comparing quasi TMDWFs as functions of $\lambda=zP^z$ of two Dirac matrices $\Gamma=\gamma^t\gamma_5$ and $\Gamma=\gamma^z\gamma_5$. In small $\lambda$ area, the behaviors of quasi TMDWFs for two Dirac matrices are a little different, which is expected to decrease with the increase of $P^z$. So the average of these two cases is likely to eliminate the power corrections. 

\newpage
\begin{widetext}

\begin{figure}
    \centering
    \includegraphics[scale=0.55]{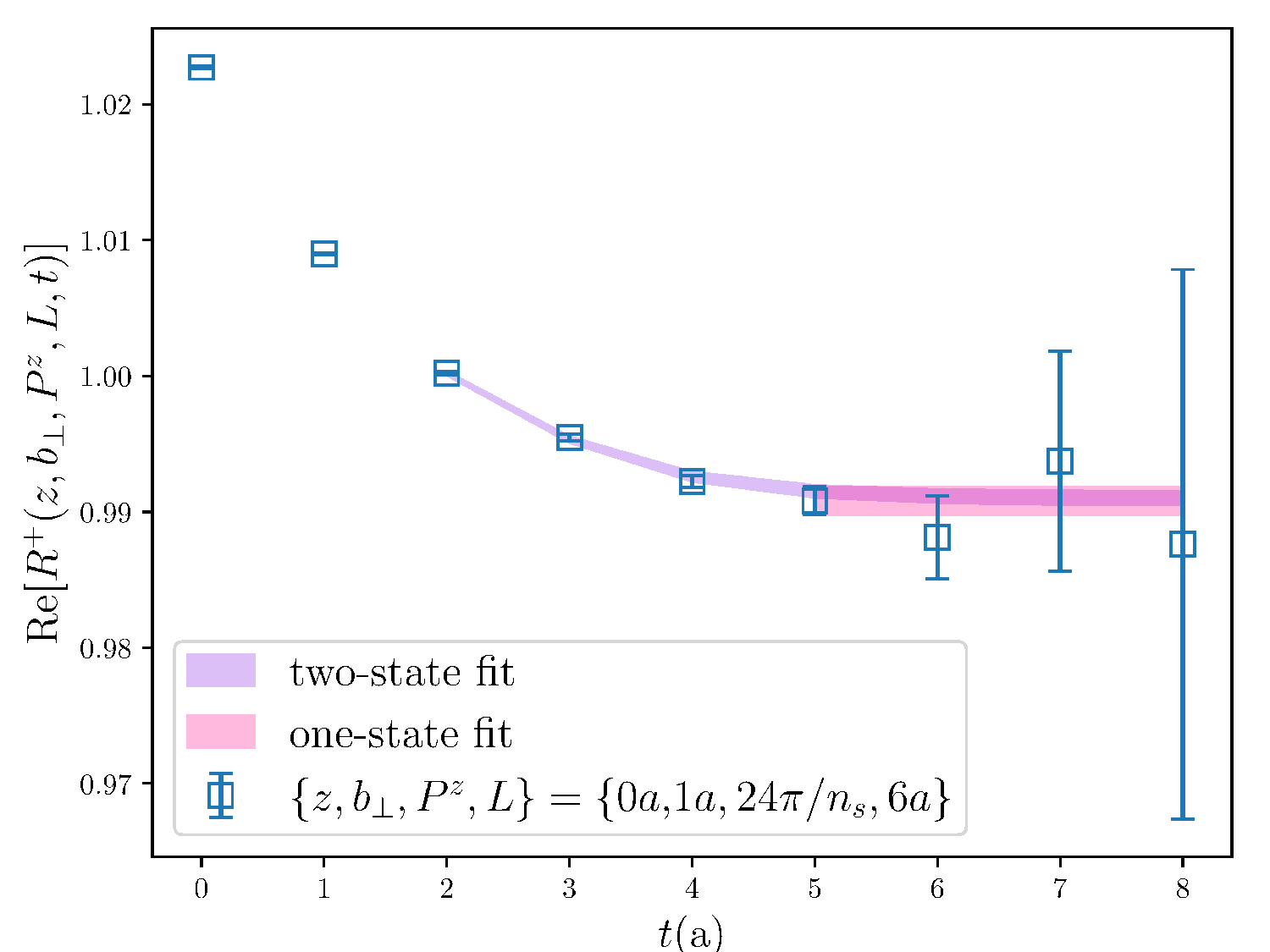}
    \includegraphics[scale=0.55]{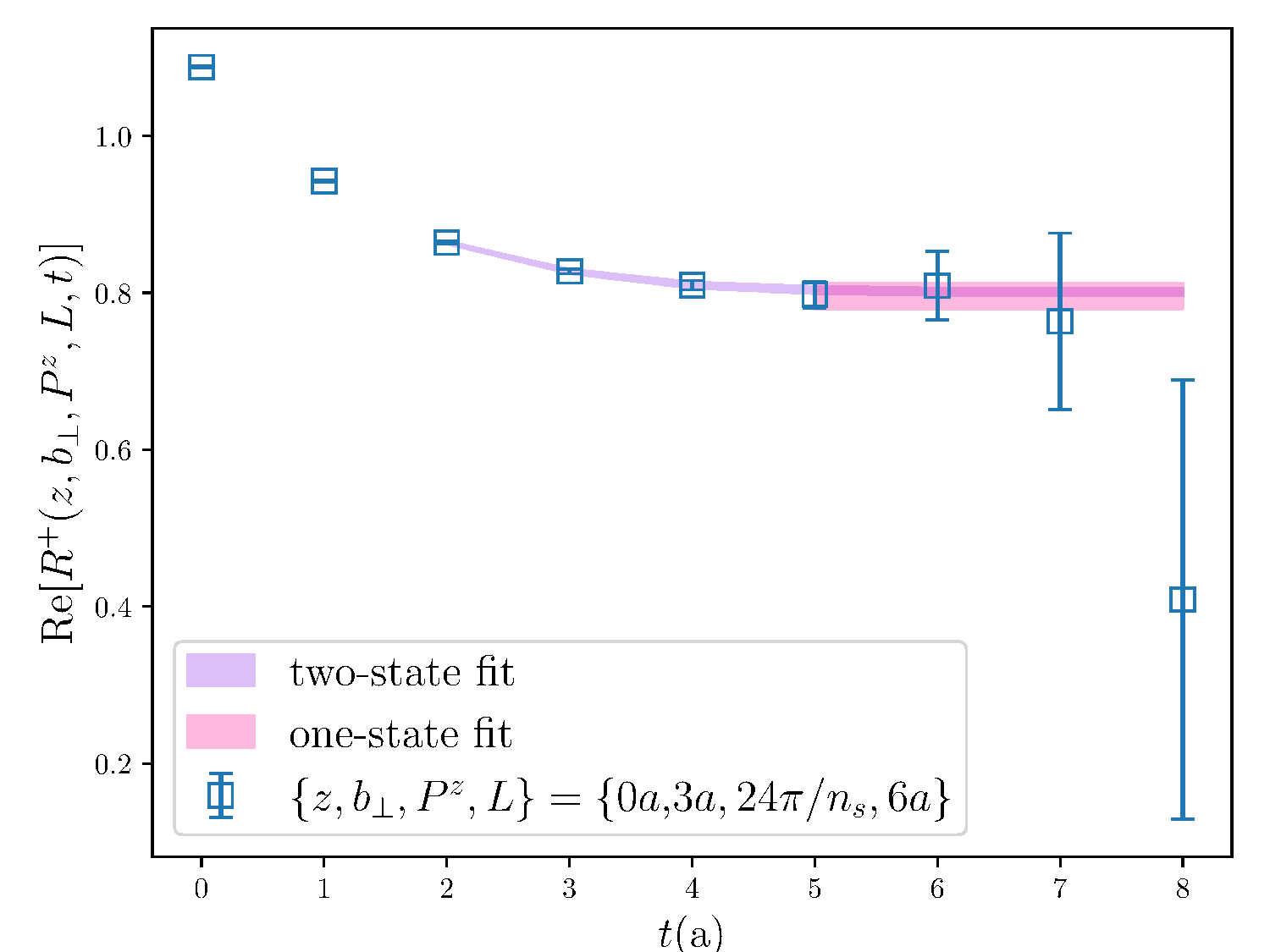}
    \includegraphics[scale=0.55]{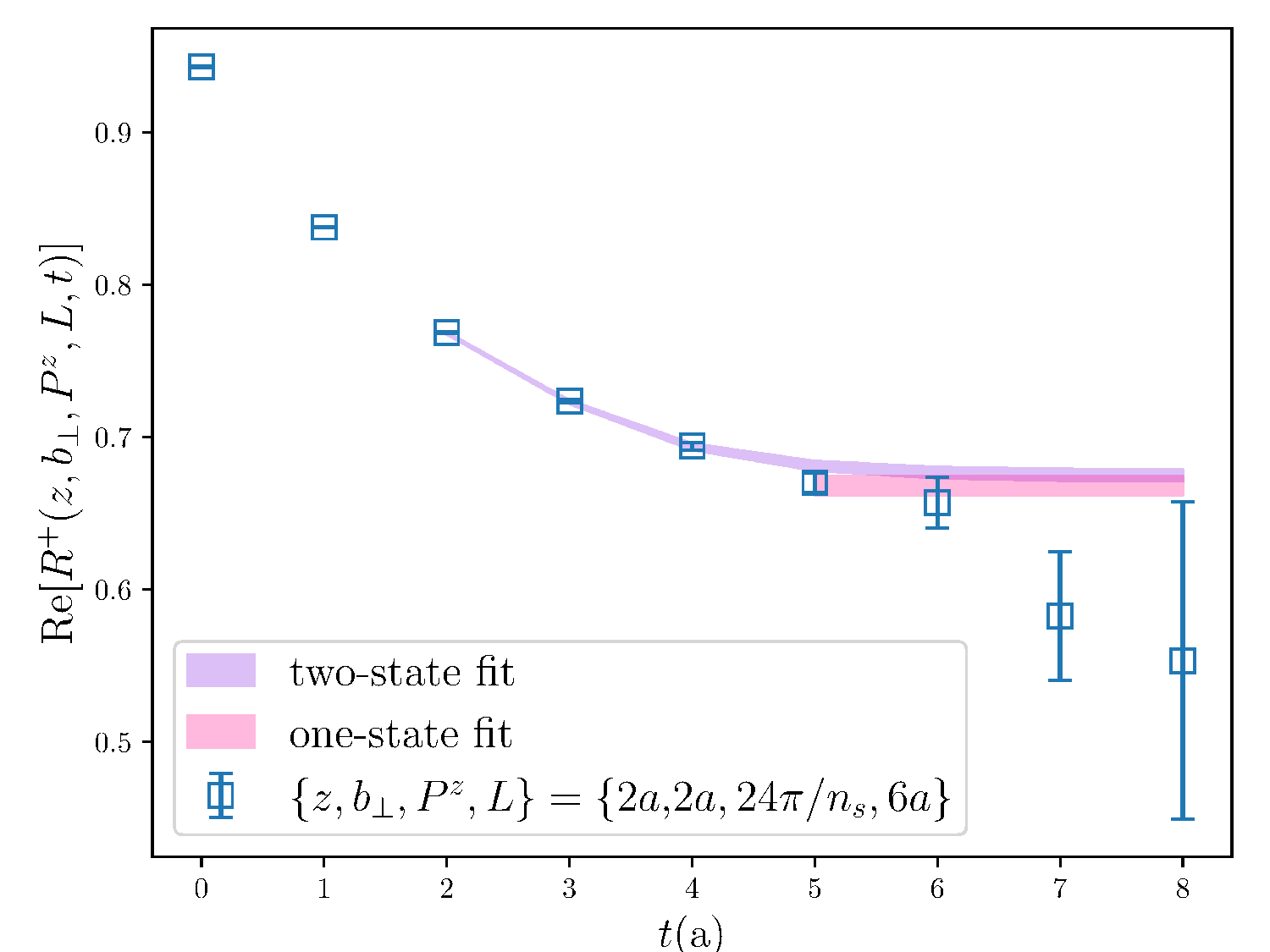}
    \includegraphics[scale=0.55]{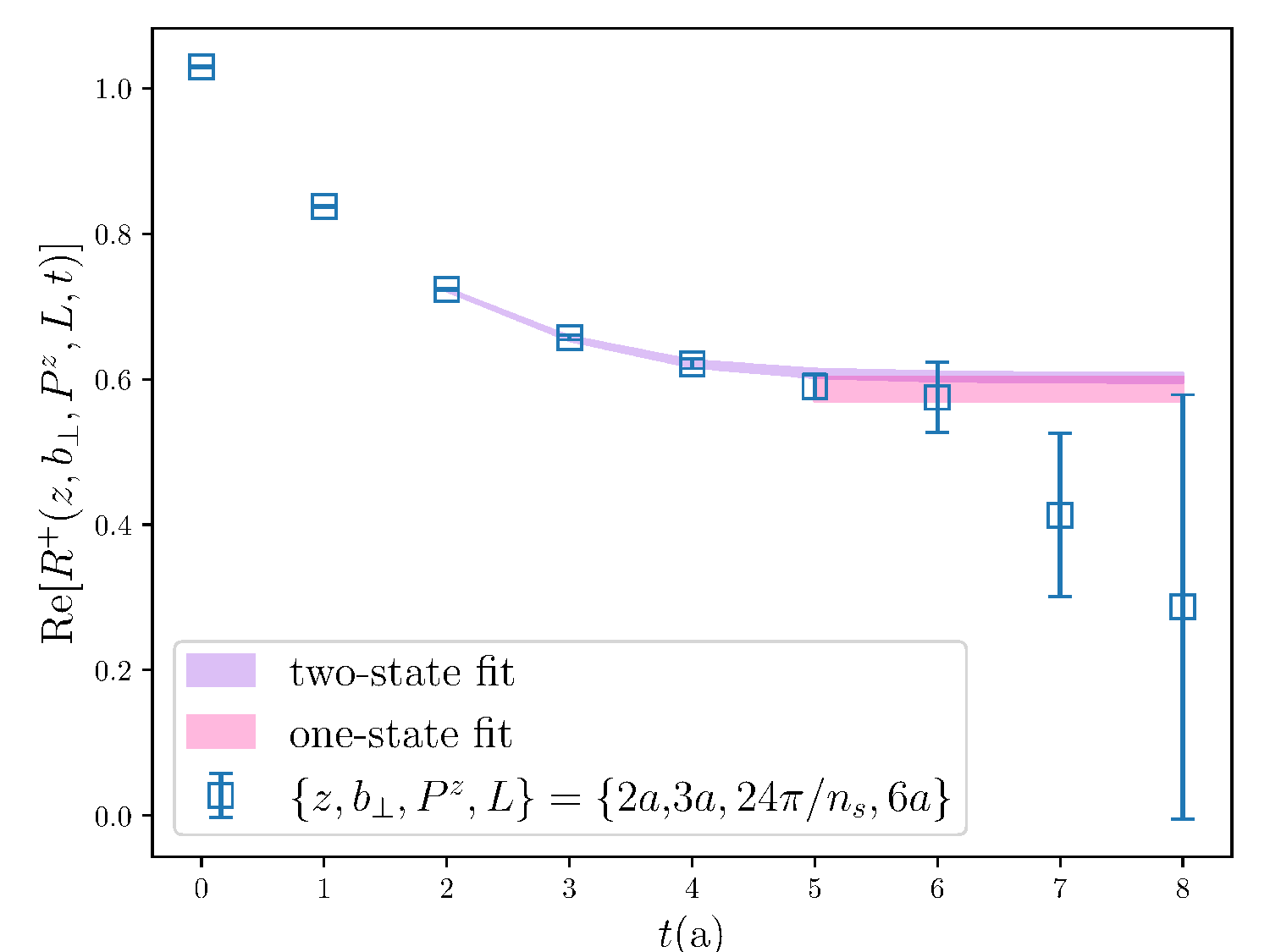}
    \caption{Four examples for comparing two-state fit and one-state fit to extract the $\tilde{\Phi}^{\pm0}(z,b_{\perp},P^z,L)$ from $R^{\pm}(z,b_{\perp},P^z,L,t)$ as described in Sec. \ref{sec:quasi-WF_from_2pt} with $\{z,b_{\perp}\}=\{0a,1a\},\{0a,3a\},\{2a,2a\},\{2a,3a\}$, and $\{P^z,L\}=\{24\pi/n_s,6a\}$. The fit range for two-state fit is $t\in[2a,8a]$, which for one-state fit is $t\in[5a,8a]$.}
    \label{fig:t_dep}
\end{figure}

\begin{figure}
    \centering
    \includegraphics[scale=0.55]{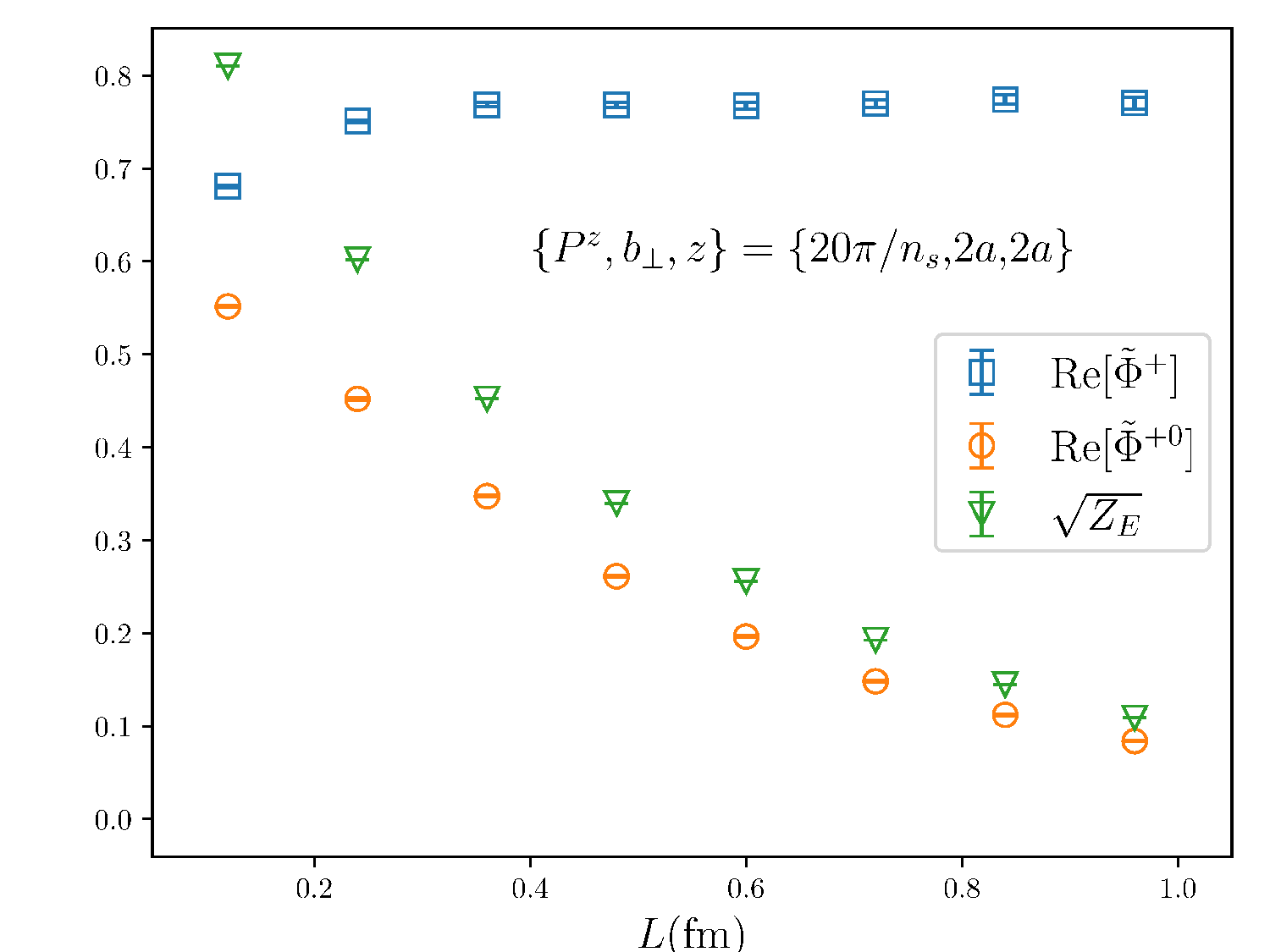}
    \includegraphics[scale=0.55]{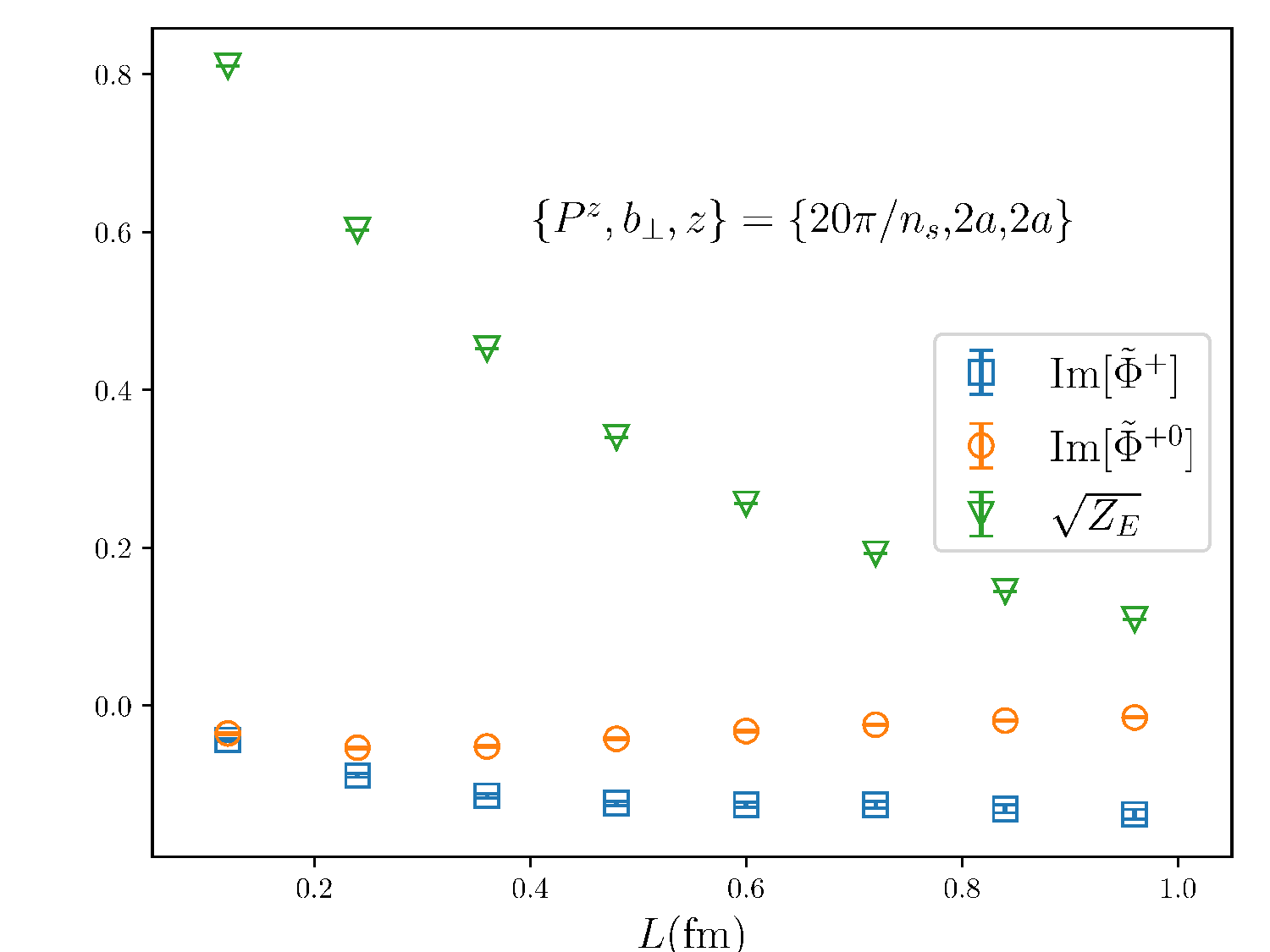}
    \includegraphics[scale=0.55]{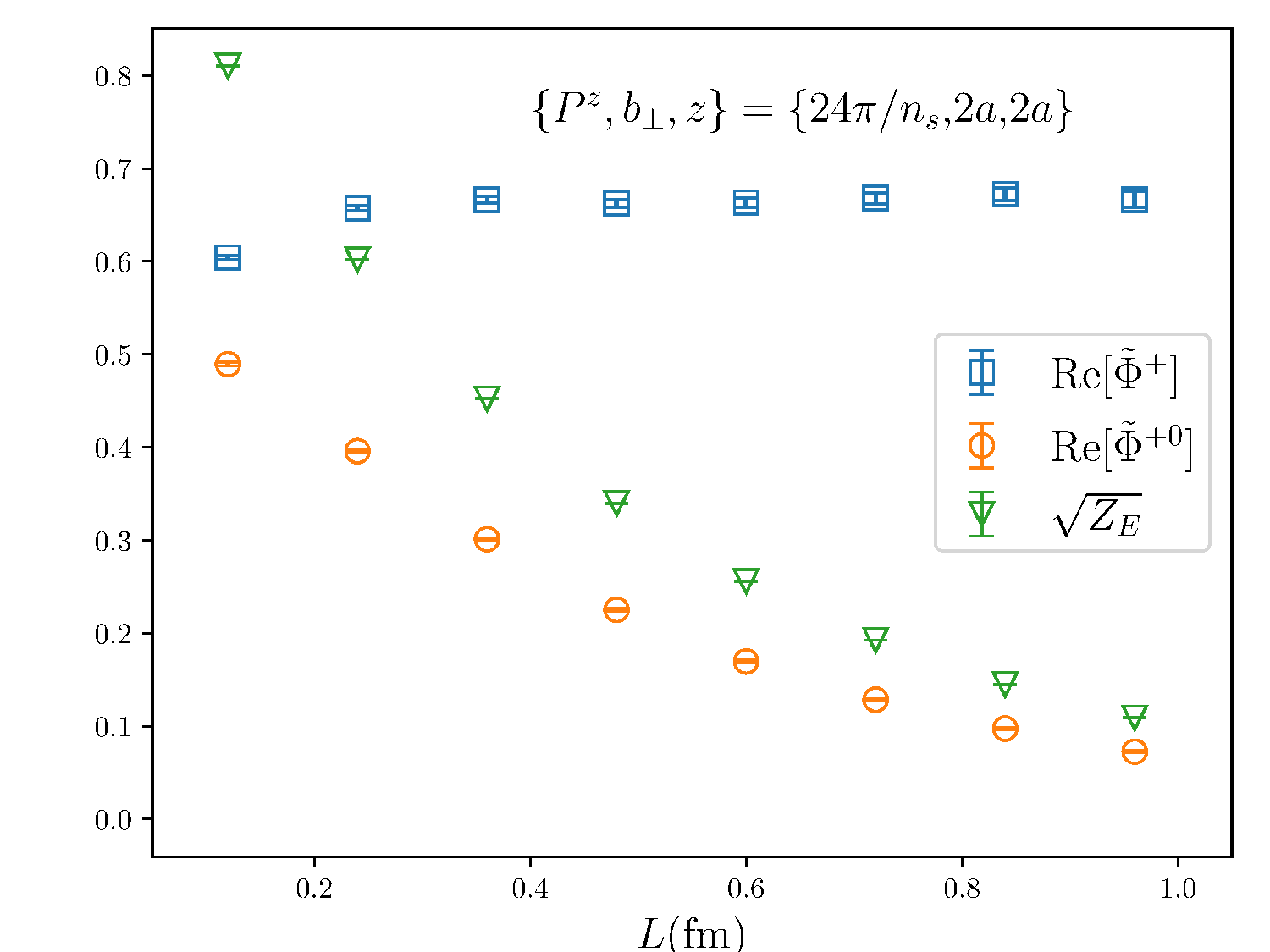}
    \includegraphics[scale=0.55]{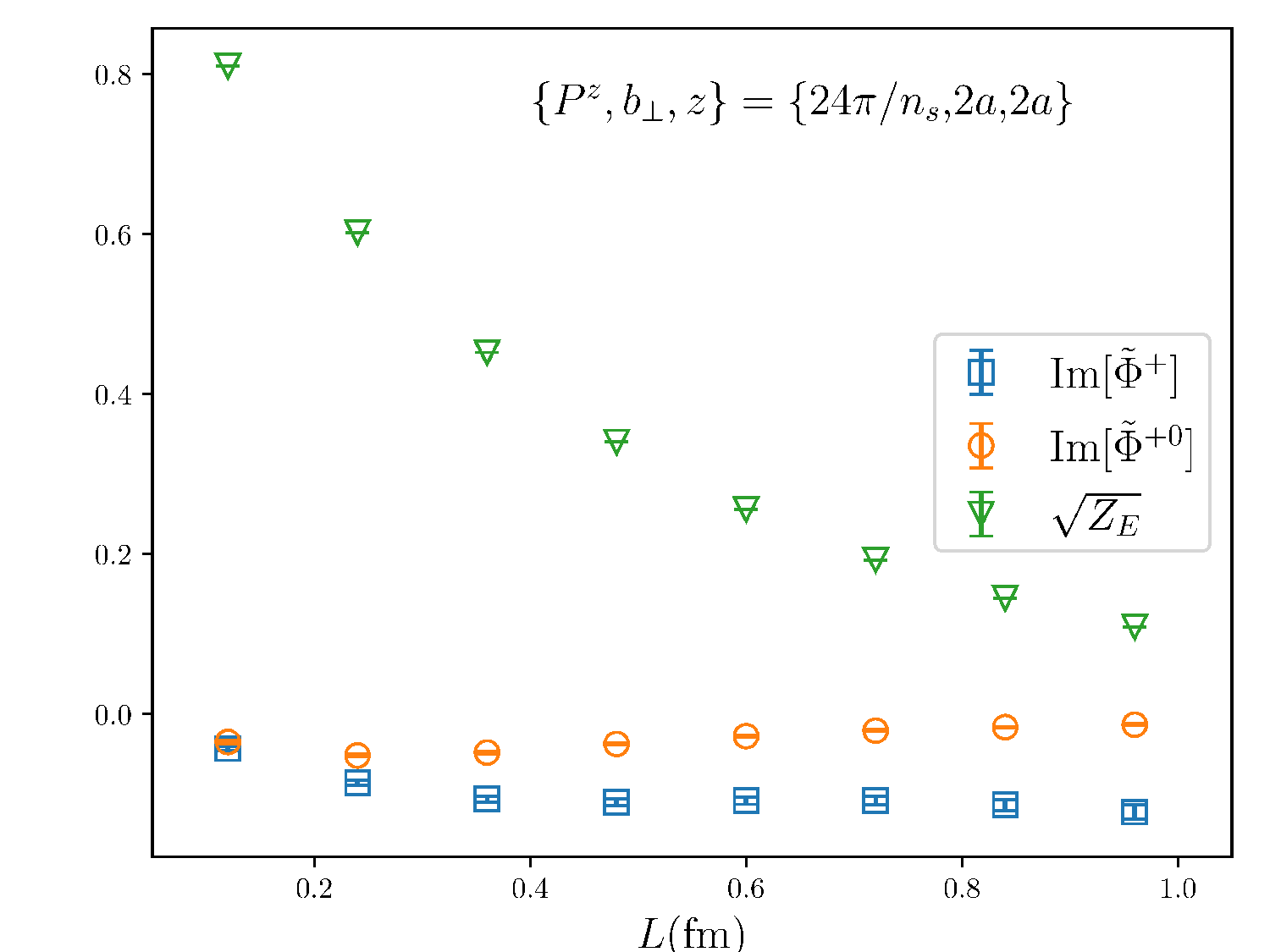}
    \includegraphics[scale=0.55]{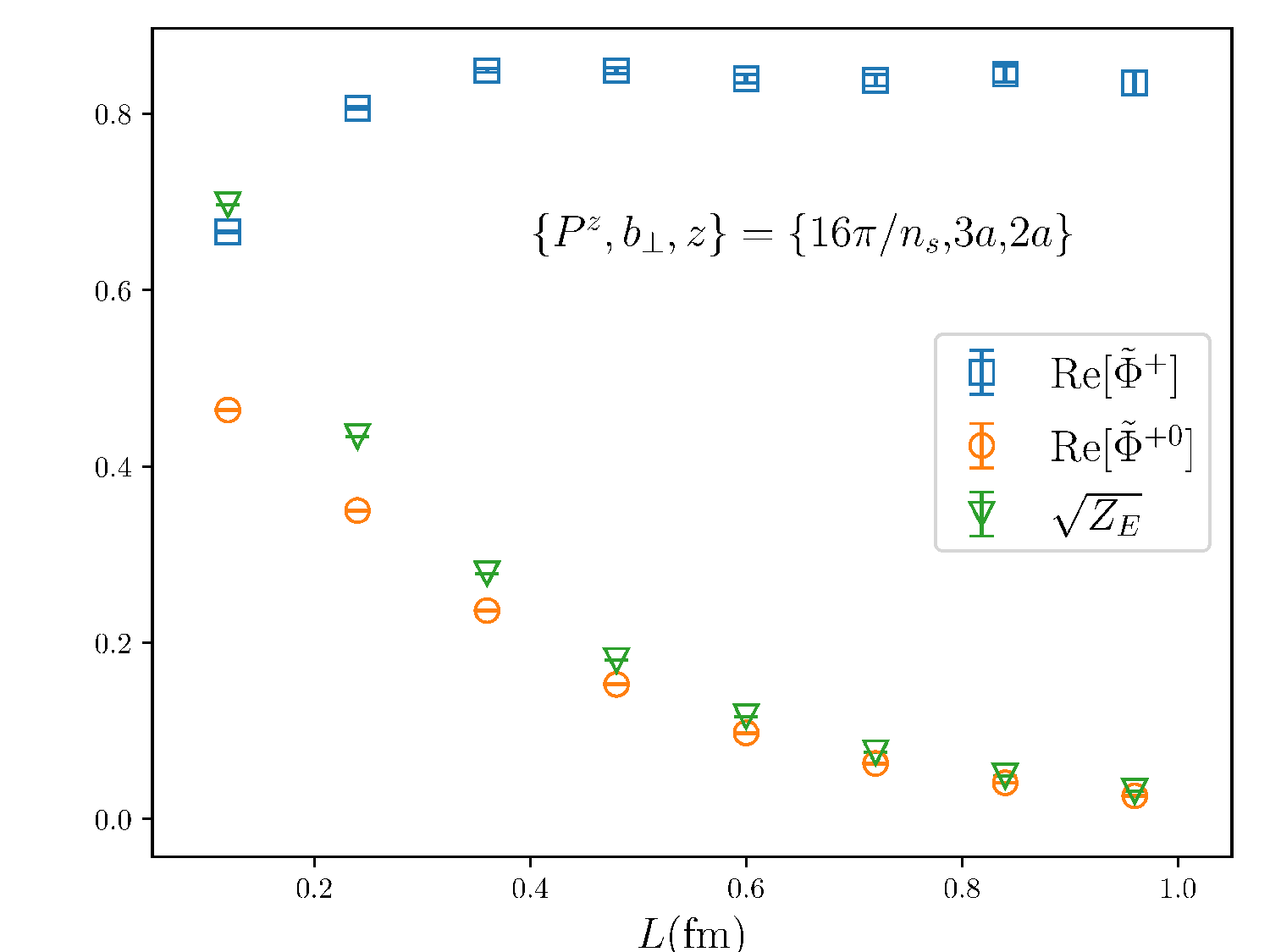}
    \includegraphics[scale=0.55]{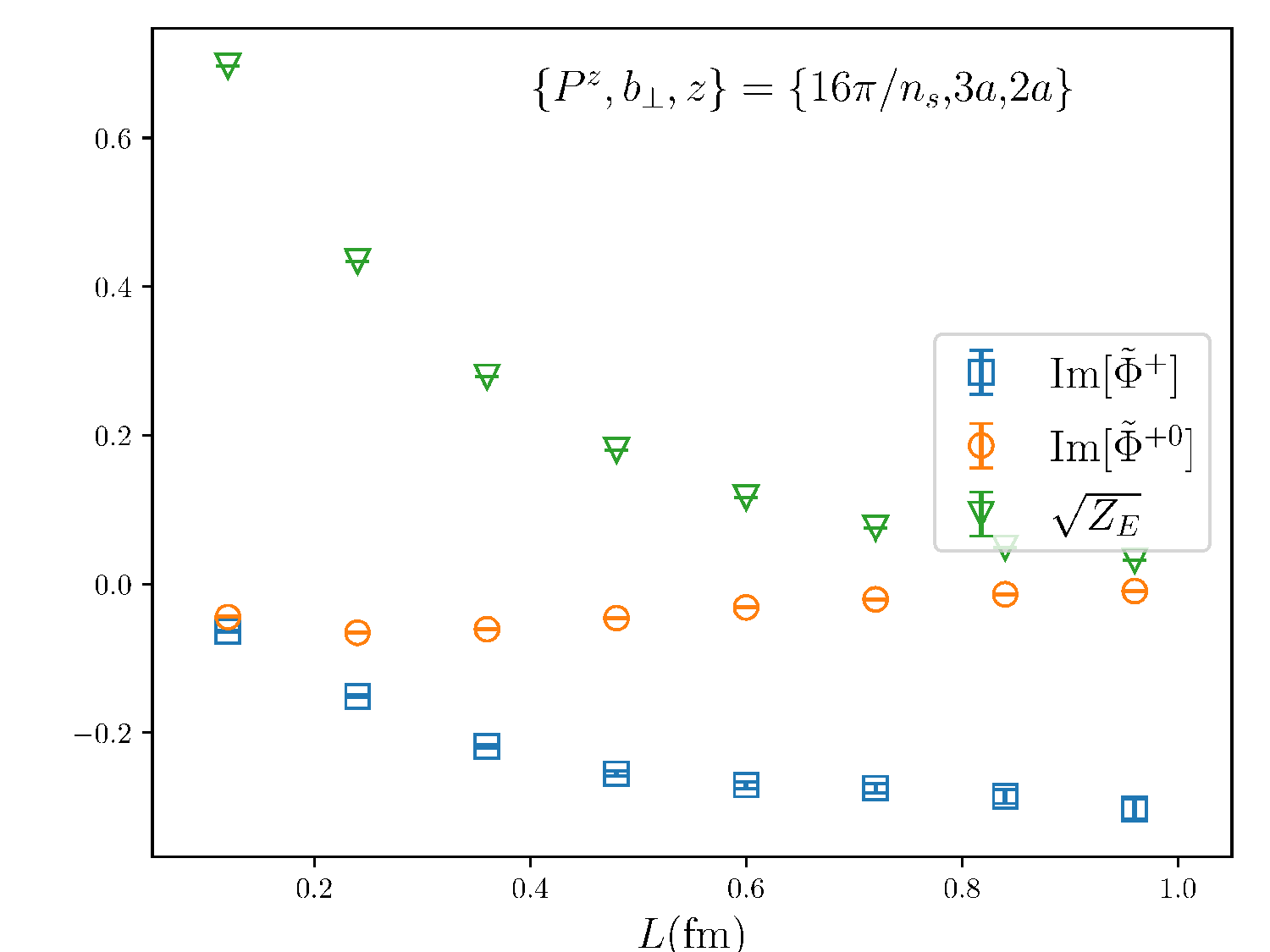}
    \caption{Results showing the dependence  on the gauge line length $L$ of unsubtracted and subtracted quasi TMDWFs as well as the Wilson loop with $\{P^z,b_{\perp},z\}$ shown in each figure. These results are for  $\Gamma=\gamma^z\gamma_5$.}
    \label{fig:multi_l_dep}
\end{figure}

\begin{figure}
    \centering
    \includegraphics[scale=0.55]{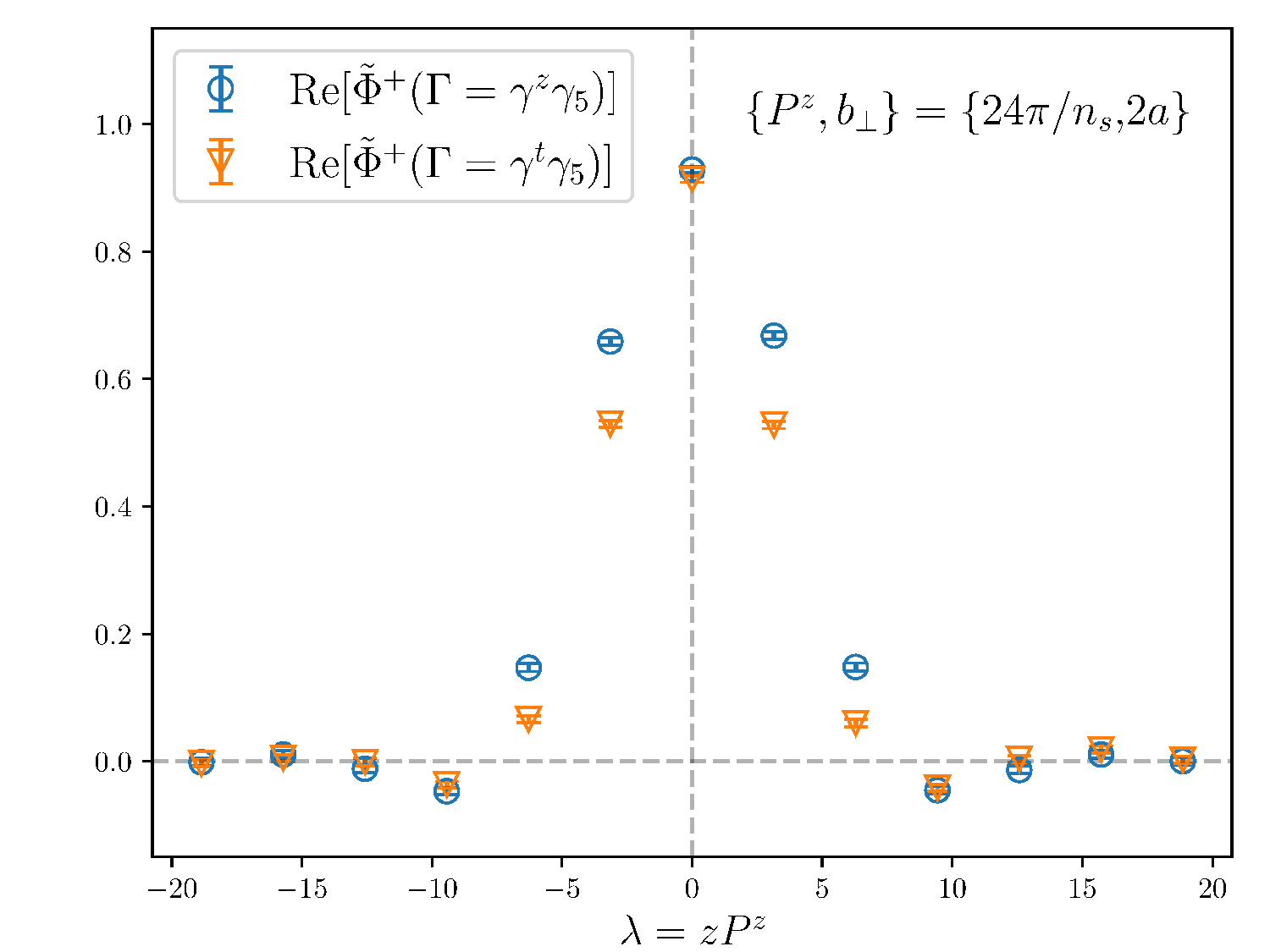}
    \includegraphics[scale=0.55]{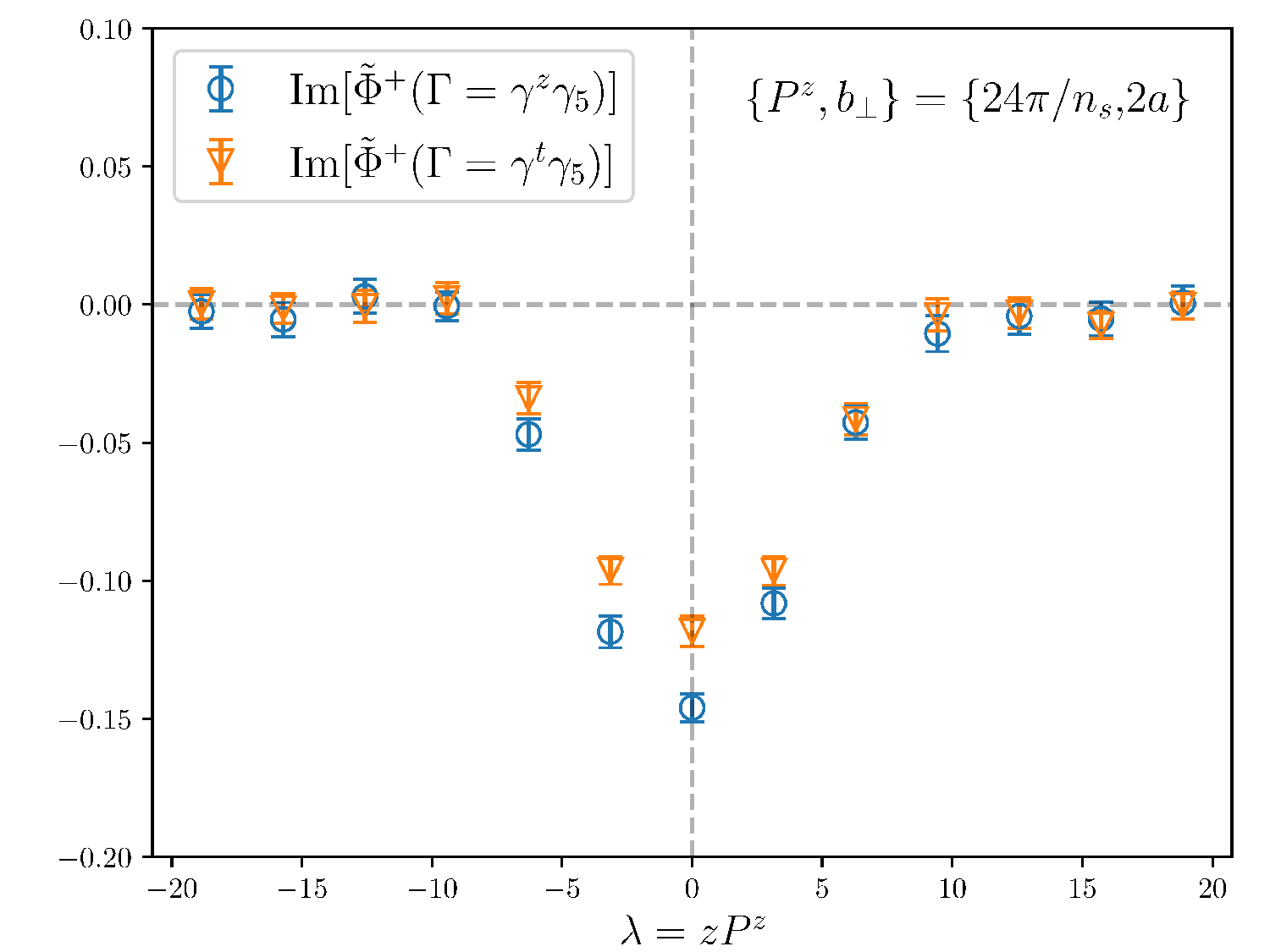}
    \includegraphics[scale=0.55]{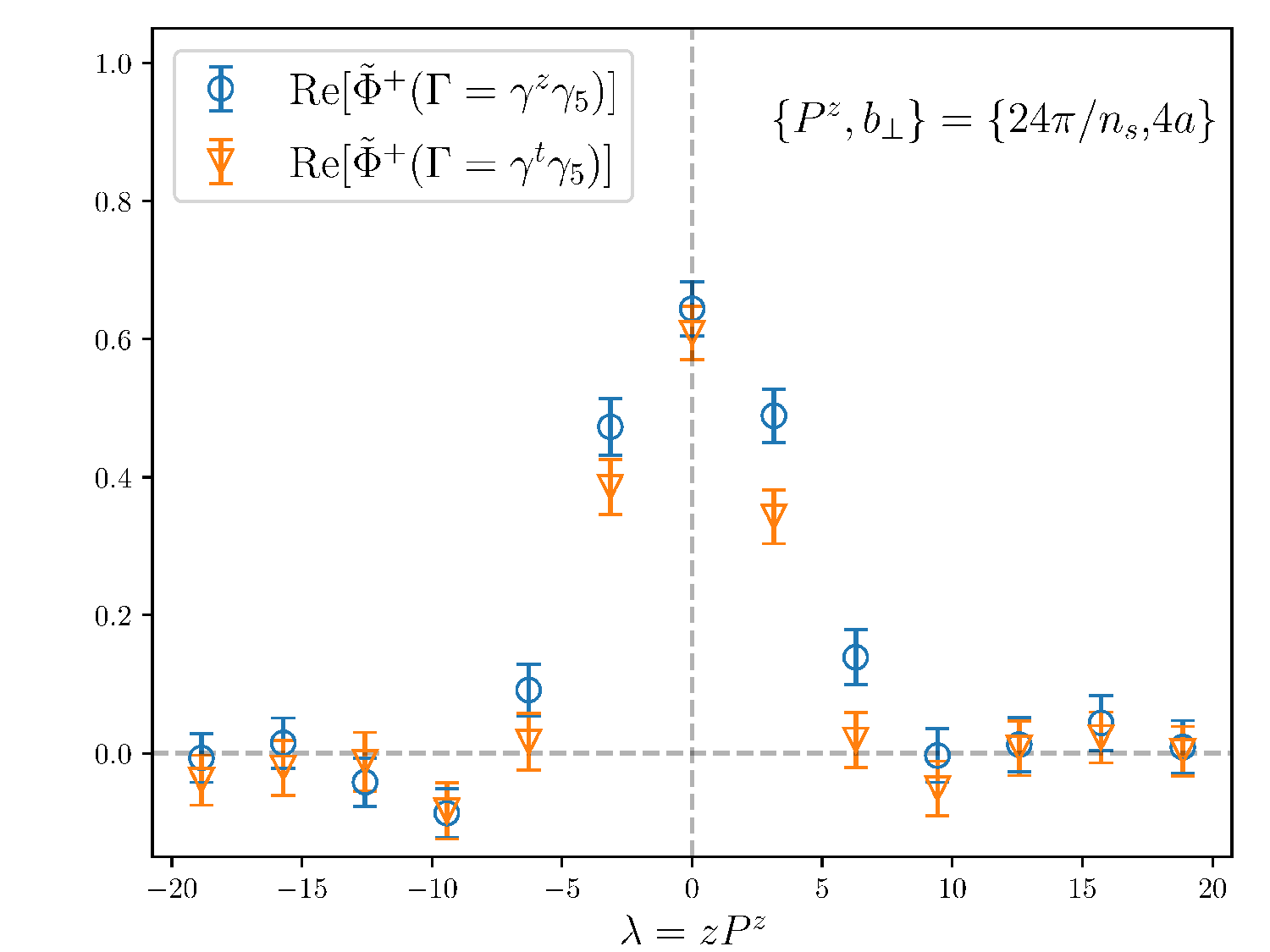}
    \includegraphics[scale=0.55]{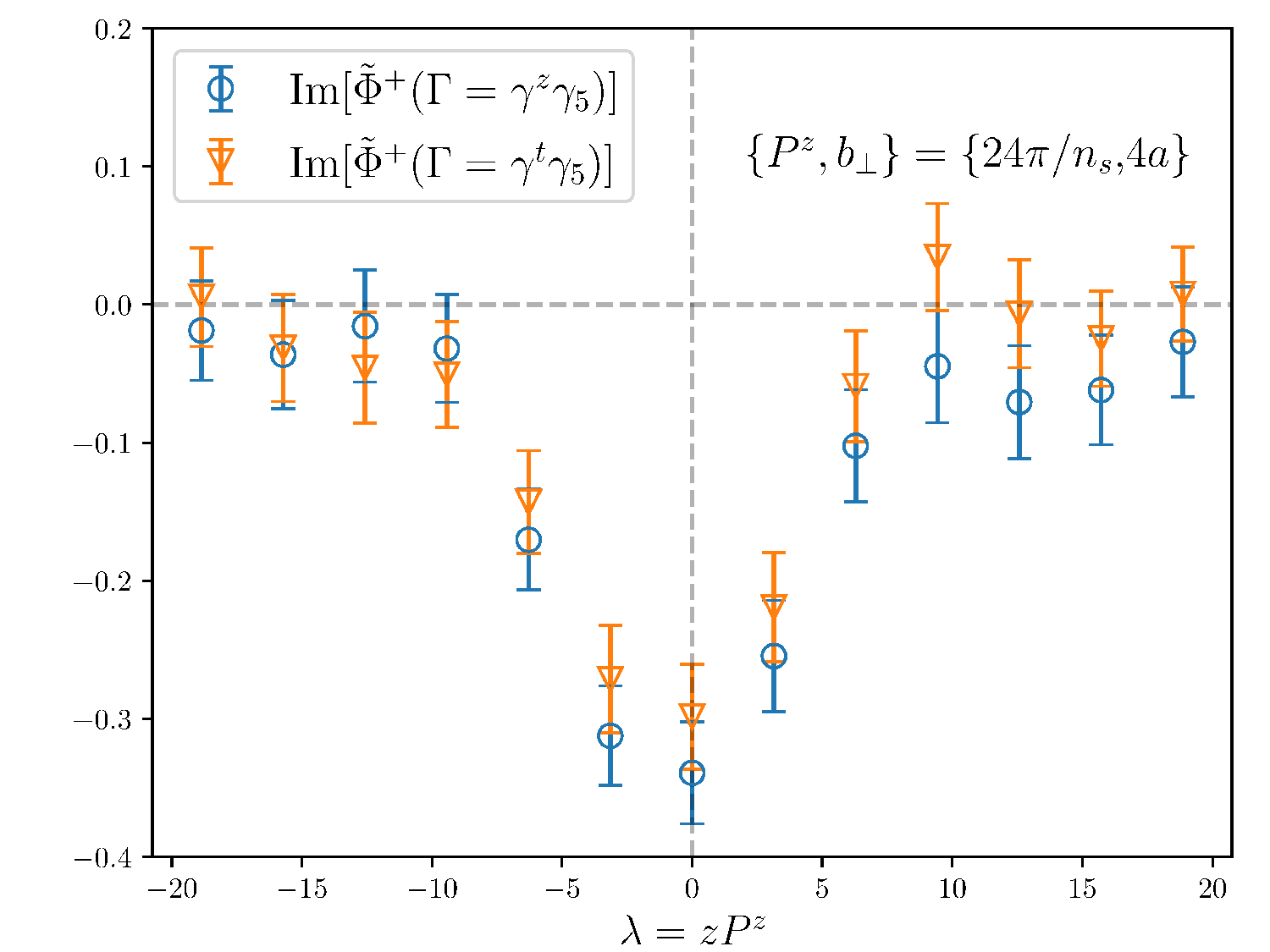}
    \includegraphics[scale=0.55]{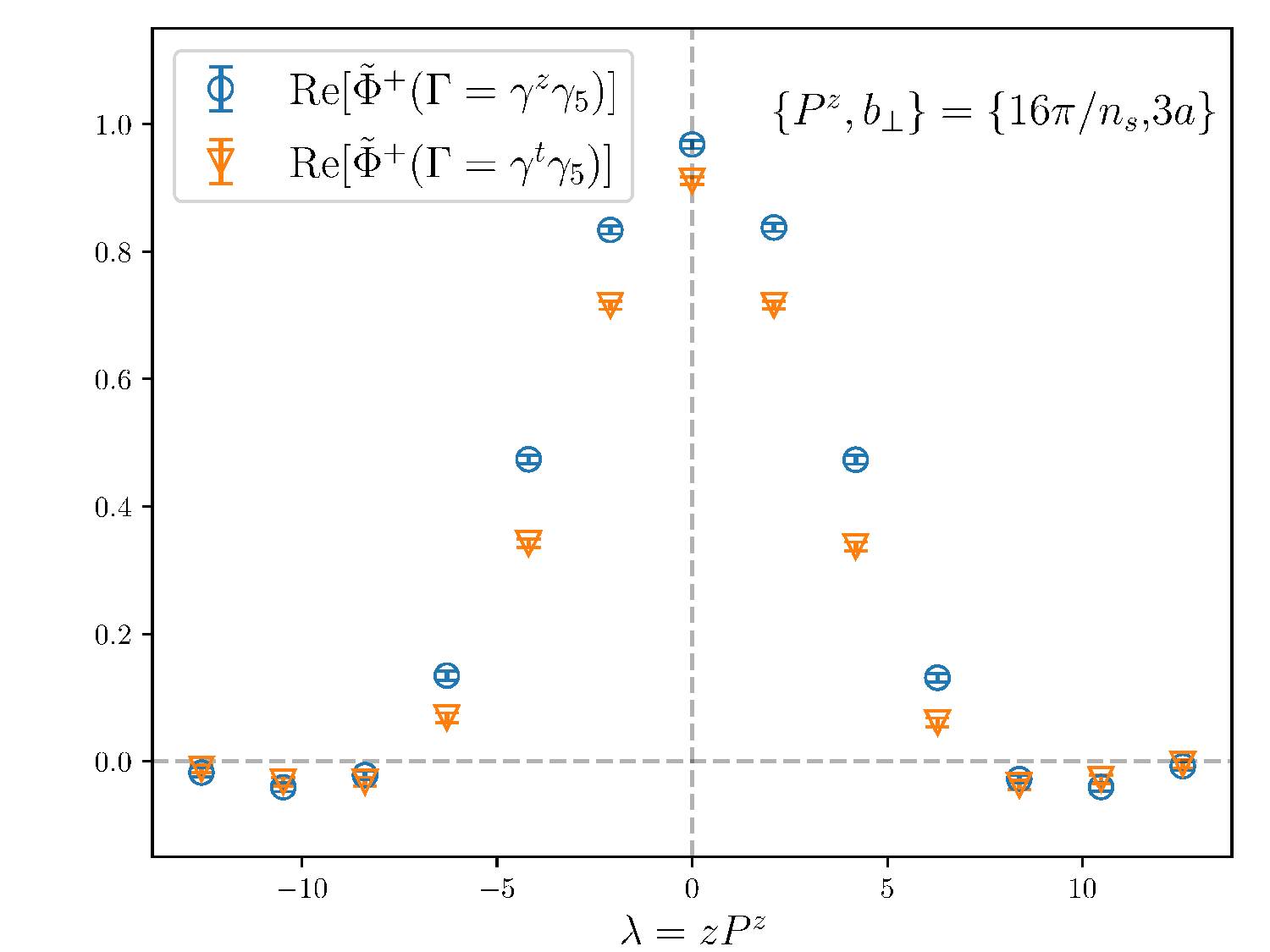}
    \includegraphics[scale=0.55]{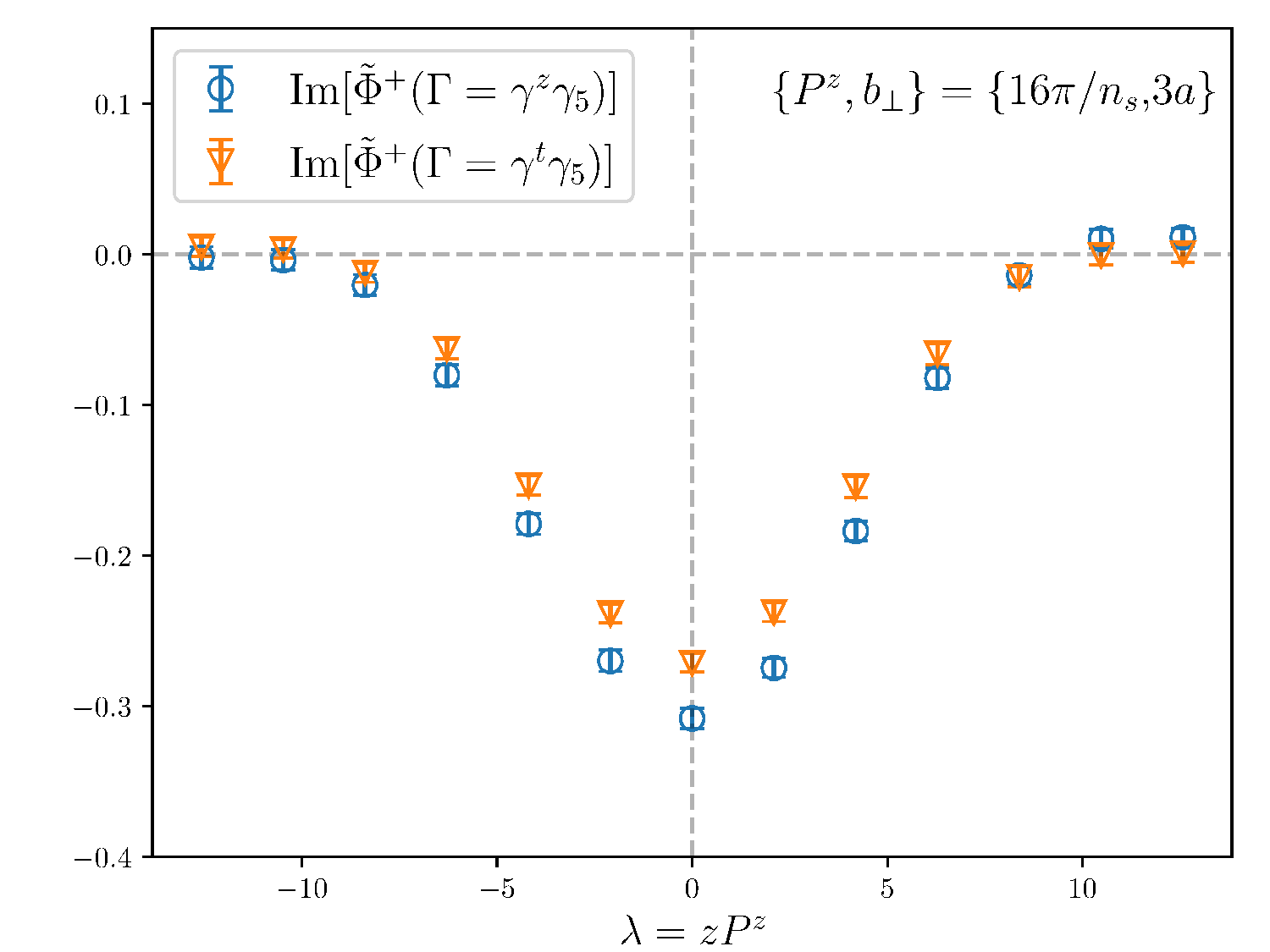}
    \caption{Examples of comparisons for $\lambda$-dependence of quasi-WF matrix elements with two Dirac matrices: $\Gamma=\gamma^t\gamma_5$ and $\Gamma=\gamma^z\gamma_5$, with $\{P^z,b_{\perp}\}$ shown in each figure. Power corrections cause the deviation between both cases.}
    \label{fig:multi_qwf_co}
\end{figure}

\end{widetext}


\bibliographystyle{unsrt}
\bibliography{main.bib}

\end{document}